\documentclass[final,journal]{IEEEtran}

\usepackage{cite}
\ifCLASSINFOpdf
\else
\fi

\usepackage{array}

\ifCLASSOPTIONcompsoc
  \usepackage[caption=false,font=normalsize,labelfont=sf,textfont=sf]{subfig}
\else
  \usepackage[caption=false,font=footnotesize]{subfig}
\fi

\usepackage[hidelinks]{hyperref}
\usepackage{url}

\usepackage{graphicx} 
\graphicspath{{./Figures/}}
\usepackage{mathtools,amsmath,amssymb,amsthm} 
\usepackage[overload]{empheq}
\usepackage{enumitem}
\usepackage{multicol}
\usepackage{capt-of}
\usepackage[export]{adjustbox}

\theoremstyle{remark}

\usepackage{siunitx}
\usepackage{booktabs}
\usepackage{xcolor}
\usepackage[ruled,vlined]{algorithm2e}
\usepackage{breqn}
\usepackage{multirow}
\usepackage{mathrsfs}
\usepackage{mathrsfs}
\usepackage{breqn}
\usepackage{mathtools}

\newcommand{\ab}{\textcolor{black}}
\newcommand{\dk}{\textcolor{black}}

\DeclareMathAlphabet\mathbfcal{OMS}{cmsy}{b}{n}

\begin{document}
%
\title{Joint Blind Deconvolution and Robust Principal Component Analysis for Blood Flow Estimation in Medical Ultrasound Imaging}
%
%
%
\author{Duong-Hung~Pham, Adrian Basarab, Ilyess Zemmoura, Jean-Pierre Remenieras and  Denis Kouam\'e    
\thanks{~D.-H. Pham, A. Basarab and D. Kouam\'e are with the IRIT Laboratoty,  Universit\'e de Toulouse, and CNRS, Toulouse 31400, France. Email:  \{duong-hung.pham; denis.kouame; adrian.basarab\}@irit.fr.}
\thanks{~Y. Zemmoura and JP. Remenieras are with the UMR 1253, iBrain, Universit\'e de Tours, Inserm, Tours, France. Email: \{ilyess.zemmoura; jean-pierre.remenieras\}@univ-tours.fr.}
}
\markboth{IEEE Transactions on Ultrasonics, Ferroelectrics, and
Frequency Control (T-UFFC)}%
{Shell \MakeLowercase{\textit{et al.}}: Bare Demo of IEEEtran.cls for IEEE Journals}
%


\maketitle


\begin{abstract}
This paper addresses the problem of high-resolution Doppler blood flow estimation from an ultrafast sequence of ultrasound images. Formulating the separation of clutter and blood components as an inverse problem has been shown in the literature to be a good alternative to spatio-temporal singular value decomposition (SVD)-based clutter filtering. In particular, \dk{a deconvolution step has recently been} embedded in such a problem to mitigate the influence of the experimentally measured point spread function (PSF) of the imaging system. Deconvolution was shown in this context to improve the accuracy of the blood flow reconstruction. However, measuring the PSF requires non-trivial experimental setups. To overcome this limitation, we propose herein a blind deconvolution method able to estimate both the blood component and the PSF from Doppler data. Numerical experiments conducted on simulated and \textit{in vivo} data demonstrate qualitatively and quantitatively the effectiveness of the proposed approach in comparison with the previous method based on experimentally measured PSF \ab{and two other state-of-the-art approaches}. 
\end{abstract}

  
\begin{IEEEkeywords}
medical ultrasound, clutter separation, blood flow, sensitive Doppler, robust PCA, blind deconvolution.
\end{IEEEkeywords}

\IEEEpeerreviewmaketitle

\section{Introduction}
\label{sec:intro}
\IEEEPARstart{T}{he} last few years have witnessed an upsurge of interest in the retrieval of high-sensitivity and high-resolution blood flow from ultrafast sequences of ultrasound (US) images. The reason behind this fact stems from the increasing demand of more accurate detection and visualization of underlying vascular structures, especially in small vessels where blood velocities become low, e.g., in cerebral or peritumoral area \cite{Mace2013} or where tissues \ab{are moving fast}, e.g., in cardiac or abdominal \ab{regions} \cite{Osmanski2012,Hur2014}. Indeed, such advances are of extreme importance to ensure a better treatment of related diseases such as brain gliomas (tumors) or peri-tumoral infiltration. To estimate the desired blood flow image, numerous methods have been proposed, whose principal goal is to suppress completely unexpected clutter signals originating from tissue components from blood flow \cite{Ashikuzzaman2020}. Among them, finite impulse response (FIR) and infinite impulse response (IIR) filters \cite{Thomas1994,Tysoe1995}, the two simplest methods, aimed to perform high-pass filtering on the ultrasonic signal along the temporal dimension. However, \ab{IIR} filters possess a long settling time whereas FIR filters need a high-order to discriminate clutter signals from blood flow. This results in an insufficient number of temporal samples at each spatial location \ab{in the case of} US images built by the means of focused ultrasonic beams \cite{Yoo2003}. Additionally, the high-pass filtering approach suffers from an intrinsic inefficiency of clutter removal when blood flow moves slowly or when tissue motion is non-negligible. Moreover, only the temporal information is taken into account for the tissue and blood separation while neglecting the importance of the high spatial coherence of the tissue compared with the blood signal. Inspired by the latter fact, a clutter reduction technique based on singular value decomposition (SVD) of the spatio-temporal (time and depth) correlation matrix of successive temporal samples was introduced in \cite{Ledoux1997}. Different techniques were subsequently proposed to extend this study such as the down-mixing approach based on an eigen-based estimation of tissue motion \cite{Bjaerum2002}, real-time eigen-based clutter rejection technique \cite{Lovstakken2006} or blood velocity estimation method using principal component analysis (PCA) \cite{Kruse2002}. However, since the use of focused beam strategies  as in conventional US imaging \ab{leads to a} low number of both temporal and spatial samples, the SVD \ab{efficiency} in all the aforementioned methods \ab{remains} limited \cite{demene_spatiotemporal_2015}.

To cope with this drawback, a recent extension of \ab{applying} SVD to the Casorati matrix of ultrafast US \ab{datasets} has been proposed in \cite{demene_spatiotemporal_2015}, and demonstrated a considerable improvement in both clutter filtering and blood flow recovery, even in the case of moving tissue and slow blood flow. Despite its efficiency, this approach strongly depends on the manual choice of two rank thresholds, used to separate the singular vectors corresponding to blood flow subspace from those corresponding to tissue and noise subspaces, thus seriously hindering its practical applicability \cite{Baranger_adaptative_2018}. To overcome this limitation, several studies have been carried out \ab{such as an efficient estimator for automatic thresholding of subspaces \cite{Baranger2018}, the use of robust principal component analysis (RPCA) for blood flow reconstruction in ultrafast US imaging \cite{Maronna_robust_1976,Wright_robust_2009,Candes2011}, or improvements of this method using either sparse regularisation in a specific basis \cite{Fatemi2018}, or sparse coding through a specific dictionary \cite{Hossack2017}}. Furthermore, embedding a deconvolution step in RPCA, called deconvolutive RPCA (DRPCA) enabled to take into account the inherent low resolution of the ultrafast Doppler data incurred by the inference of the point spread function (PSF) characterizing the imaging system in use \cite{Shen2019}. However, DRPCA is based on the knowledge of the PSF that requires to be measured by an independent acquisition procedure, thereby causing considerable inconvenience to users or being infeasible in \dk{practical} situations.  

In an independent work \cite{Michailovich2019}, Michailovich et \emph{al.} introduced an appealing method for blind deconvolution (BD) of 2D US images. This method enabled an accurate PSF estimation together with the reconstruction process of a 2D US image with an increased spatial resolution and improved contrast, using only partial information of the PSF, i.e., its power spectrum. \ab{Investigating} how the joint analysis of both DRPCA and BD methods can be beneficial for \ab{joint} blood flow retrieval and PSF estimation in the \ab{context of} ultrafast US imaging \ab{represents} the main goal of the present paper. The remainder of the paper is organized as follows. Basics about SVD, DRPCA and BD are regrouped in Section \ref{sec:estab}. The proposed algorithm of blood flow estimation is detailed in Section \ref{sec:BDRPCA}. Finally, results on both simulated and \textit{in vivo} ultrafast US data,  demonstrating the \dk{effectiveness of the proposed approach  over state-of-the-art methods}, are reported in Section \ref{sec:Results}.


\section{Background}\label{sec:estab}
In this section, nomenclature and notations related to SVD, DRPCA and BD methods are provided. Throughout this paper, $z$ denotes a scalar, $\boldsymbol{z}$  a vector and $\boldsymbol{Z}$ a matrix. Subscript $\boldsymbol{Z}_{n,m}$ denotes the element in the $n$th row and $m$th column of $\boldsymbol{Z}$ while superscript $\boldsymbol{Z}^{(k)}$ represents $\boldsymbol{Z}$ at iteration $k$. In some contexts, the vectorized counterpart of the matrix $\boldsymbol{Z}$ \ab{in standard lexicographical order} will be considered, defined by the same notation $\boldsymbol{Z}$. 

\subsection{Problem formulation}
Let us \ab{consider the value of} a complex analytic signal after demodulation at a given lateral and axial coordinates $(x,z)$ \ab{at time, i.e., frame number,} $t$, \ab{written as:}
\begin{equation*}
S(x,z,t) = I(x,z,t) + i Q(x,z,t),
\end{equation*} 
where $I(x,z,t)$ and $Q(x,z,t)$ are the in-phase and quadrature (IQ) components. This IQ signal is commonly modelled as a superposition of the tissue signal $T(x,z,t)$, the blood  signal $B(x,z,t)$ and \dk {an} additive noise component $N(x,z,t)$ as follows \cite{demene_spatiotemporal_2015}:
  \begin{equation*}
S(x,z,t) = T(x,z,t) + B(x,z,t)+ N(x,z,t).
\end{equation*} 
\ab{Assume} that $N_t$ RF frames of size $N_z \times N_x$, \ab{with $N_z$ the dimension in the axial direction and $N_x$ the dimension in the lateral direction,} are recorded via ultrafast US imaging. \ab{Constructing the Casorati matrix from this 3D Doppler data}, i.e. stacking \ab{these} frames into a 2D matrix, leads to the  following model:    
\begin{equation}
\boldsymbol{S} = \boldsymbol{T}+\boldsymbol{B}+\boldsymbol{N},
\label{eq:basic}
\end{equation}
\noindent where $\boldsymbol{S}, \boldsymbol{T}, \boldsymbol{B}$  and $\boldsymbol{N}  \in \mathbb{C}^{N_z N_x\times N_t}$ are respectively the dataset Casorati, tissue, blood and noise matrices. \ab{The} main goal \ab{of this work} is to formulate an optimization problem to able to retrieve $\boldsymbol{B}$ and $\boldsymbol{T}$ from $\boldsymbol{S}$ under some constraints \ab{imposed to} these matrices. Before going into \ab{the details of the proposed method}, the following \ab{subsections} \dk{summarize} the most \dk{common} \ab{methods for blood flow retrieval based on SVD and RPCA}.     

\subsection{SVD-based method}
\label{subsec:SVD}
This method consists in factorizing the Casorati matrix $\boldsymbol{S}$ by the means of SVD as follows \cite{demene_spatiotemporal_2015}:
\begin{equation*}
\boldsymbol{S} = \boldsymbol{U} \boldsymbol{\Sigma} \boldsymbol{V}^{\dagger} = \sum\limits_{k=1}^{r}\boldsymbol{u}_{k} \sigma_{k} \boldsymbol{v}_{k}^{\dagger}, 
\end{equation*}
where $\boldsymbol{U} \in \mathbb{C}^{N_z N_x\times N_z N_x}$ and $\boldsymbol{V} \in \mathbb{C}^{N_t \times N_t}$ are respectively unitary matrices consisting of the spatial (left) $\boldsymbol{u}_{k}$ and temporal (right) $\boldsymbol{v}_{k}$ singular vectors \ab{and} $\boldsymbol{\Sigma} \in \mathbb{R}^{N_z N_x\times N_t} $ is a non-square diagonal matrix whose entries are \ab{its} singular values $\sigma_{k}$. The superscript ${\dagger}$ stands for the conjugate transpose while $r = \min(N_z N_x, N_t)$ denotes the \ab{rank of $\boldsymbol{S}$}. Note that the $\boldsymbol{\Sigma}$ entries are sorted in a descending order such that the first largest singular values are associated with the tissue signal with the \ab{highest} energy and spatial coherence, the \ab{intermediate} ones with the blood signal and the \ab{smallest ones} with the noise. Besides, since $N_z N_x\times N_z N_x$ is generally a very large number, in practice only the first $N_z N_x\times N_t$ elements of $\boldsymbol{U}$ and $N_t \times N_t$ elements of $\boldsymbol{\Sigma}$ are computed to reduce running time and memory burden. After \ab{having} determined the clutter $T_c$ and blood $T_b$ rank thresholds, \ab{generally} by manual tuning, to remove the contribution of the tissue signal and the noise, the blood flow component is estimated by: 
\begin{equation}
\label{eq:B_SVD}
\boldsymbol{\hat{B}} =  \sum\limits_{k=T_c}^{T_b}\boldsymbol{u}_{k} \sigma_{k} \boldsymbol{v}_{k}^{\dagger}.
\end{equation}   

\subsection{RPCA-based methods}
\label{subsec:RPCA}
\ab{RPCA is an interesting alternative to SVD, able to explicitely take into account prior knowledge about the blood flow. Specifically, the blood flow $\boldsymbol{B}$ can be assumed to be sparse in number of practical applications}. Conventionally, \ab{sparsity is} promoted by the \ab{minimization of the} $l_1$-norm. \ab{In contrast to the blood component, tissue $\boldsymbol{T}$ possesses a very small change over time and can be thus considered \dk{as having a low rank}, usually modelled by the nuclear norm denoted by $||.||_*$}. Assuming  $\boldsymbol{N}$ is a Gaussian noise and taking into account these two assumptions, RPCA is expressed as the following \dk{optimization} problem:  
{\small
\begin{align}
[\hat{\boldsymbol{B}},\hat{\boldsymbol{T}]} =
\mathop{\arg\min_{\boldsymbol{B},\boldsymbol{T}}} \left\{ ||\boldsymbol{S}-\boldsymbol{B}-\boldsymbol{T}||^2_F +
\lambda||\boldsymbol{B}||_1 + \rho ||\boldsymbol{T}||_* \right\},
\label{eq:rpca_con}
\end{align}}
\noindent where $\hat{.}$ denotes the estimated variables, $||.||_F$ is the Frobenius norm and $\lambda, \rho>0$ are two hyperparameters balancing the trade-off between the sparsity of the blood and the low-rankness of the tissues \cite{Wright_robust_2009,Shen2019}. 

A common technique to solve the above problem is to use the augmented Lagrangian-based alternating direction method of multipliers (ADMM), by solving iteratively several sub-problems over each variable separately \cite{Boyd_distributed_2010}. The augmented Lagrangian related to \eqref{eq:rpca_con} can be written as follows: 
\begin{equation*}
\mathscr{L}(\boldsymbol{B},\boldsymbol{T},\boldsymbol{\nu}) = \lambda||\boldsymbol{B}||_1 + \rho||\boldsymbol{T}||_* + \frac{\mu}{2}||\boldsymbol{S}-\boldsymbol{B}-\boldsymbol{T}+\dk{\frac{1}{\mu}\boldsymbol{\nu}||^2_F},
\label{eq:rpca_lagran}
\end{equation*}
where $\boldsymbol{\nu}$ is the Lagrange multiplier and $\mu$ is \dk{the Lagrangian penalty parameter} controlling the convergence speed of the algorithm. At each iteration $k$, ADMM performs the following three steps, until a predefined \emph{stopping} criteria is met:
{\small
\begin{align}
  &\hat{\boldsymbol{B}}^{(k+1)}=\mathop{\arg\min_{\boldsymbol{{B}}}}(\lambda||\boldsymbol{B}||_1+\dk{\frac{\mu}{2}||\boldsymbol{B}-(\boldsymbol{S}-\boldsymbol{T}^{(k)}+\frac{1}{\mu} \boldsymbol{\nu}^{(k)})||^2_F}) \notag \\
   &\hat{\boldsymbol{T}}^{(k+1)}=\mathop{\arg\min_{\boldsymbol{{T}}}}(\rho||\boldsymbol{T}||_*+\dk{\frac{\mu}{2}||\boldsymbol{T}-(\boldsymbol{S}-\boldsymbol{B}^{(k+1)}+\frac{1}{\mu} \boldsymbol{\nu}^{(k)})||^2_F}) \notag \\
   &\boldsymbol{\nu}^{(k+1)}=\dk{\boldsymbol{\nu}^k+\mu(\boldsymbol{S}-\boldsymbol{B}^{(k+1)}-\boldsymbol{T}^{(k+1)})}. \notag
\end{align}
}

It is worth noting that the two first steps above are convex problems possessing closed-form solutions: soft thresholding (ST)\cite{Lin_augmented_2009} and singular value thresholding (SVT) \cite{Cai2010}, respectively.

Although powerful, the standard RPCA approach does not consider the significant impact of the PSF limiting the spatial resolution of the Doppler data. To deal with \ab{this challenge}, a deconvolution step using a measured PSF was combined with the RPCA model, resulting into the DRPCA algorithm in \cite{Shen2019}. Within this method, the blood flow $\boldsymbol{B}$ is modelled as: 
\begin{equation*} \label{eq:cirHx1}
\boldsymbol{B} = \mathbfcal{H}_m\circledast \boldsymbol{X},
\end{equation*}
where $\circledast$ stands for the 2D convolution, $\mathbfcal{H}_m$ is the measured PSF and $\boldsymbol{X} \in \mathbb{C}^{N_z N_x\times N_t}$ is the high resolution blood component to be estimated. It is worth mentioning here that using circulant boundary conditions for computational efficiency \ab{in the Fourier domain} and considering the vectorized version of $\boldsymbol{X}$, such a 2D convolution product can be written as: 
\begin{equation*} \label{eq:cirHx2}
\boldsymbol{B} = \boldsymbol{H}\boldsymbol{X},
\end{equation*}
where $\boldsymbol{H} \in \mathbb{C}^{N_z N_x N_t \times N_z N_x N_t}$ is a block circulant with circulant blocks (BCCB) matrix obtained from the PSF $\mathbfcal{H}_m$ (see, e.g., \cite{Hansen2006}). Plugging this convolution model in \eqref{eq:rpca_con} results into: 
{ 
\begin{align}
[\hat{\boldsymbol{X}},\hat{\boldsymbol{T}}] &=
\mathop{\arg\min_{{\boldsymbol{x}},{\boldsymbol{T}}}} \left\{ ||\boldsymbol{S}-\boldsymbol{H} \boldsymbol{X}-\boldsymbol{T}||^2_F \right. \notag\\
      &\hspace{3.8cm} + \left. \lambda||\boldsymbol{X}||_1 + \rho||\boldsymbol{T}||_* \right\}.
\label{eq:drpca_2}
\end{align}}
Similar to \eqref{eq:rpca_con}, \eqref{eq:drpca_2} can be also solved using an ADMM-based algorithm, as shown in \cite{Shen2019}. \ab{However}, DRPCA requires the knowledge of the PSF $\mathbfcal{H}_m$. In \cite{Shen2019}, it was measured experimentally, using the same imaging system as the one used for \textit{in vivo} data acquisition, but on a dedicated wire phantom. \ab{As highlighted previously in the introduction, the need of experimentally measuring the PSF is an important limitation of DRPCA approach}.

\subsection{Blind deconvolution} \label{subsec:BD}

Blind deconvolution (BD) methods aim at retrieving \ab{a} high resolution image from its corresponding low resolution observed image, modelled as the 2D convolution of \dk{the  high resolution image and a poorly determined or unknown PSF}. In US imaging, BD has been recently used to estimate both the spatially invariant PSF and the tissue reflectivity function (TRF) from 2D RF images \cite{Michailovich2019}. Concretely, considering a 2D IQ image $\boldsymbol{G}$ is the low resolution version of the TRF $\boldsymbol{F}$ degraded by a PSF $\mathbfcal{H}_e$ and a noise $\boldsymbol{\Omega}$, it can be written as $\boldsymbol{G} = \mathbfcal{H}_e \circledast \boldsymbol{F} + \boldsymbol{\Omega}$. \ab{Note that subscript $e$ denotes the PSF $\mathbfcal{H}_e$ to be \textit{estimated}, in contrast to the measured PSF $\mathbfcal{H}_m$ defined in the previous subsection}. \ab{Following \cite{Michailovich2019}}, BD \ab{in US imaging} can be formulated as the following optimization problem:    
\begin{align}
&[\hat{\mathbfcal{H}_e},\hat{\boldsymbol{F}}]=\mathop{\arg\min_{{\mathbfcal{H}_e},{\boldsymbol{F}}}} \left\{\frac{1}{2}||\boldsymbol{G}-\mathbfcal{H}_e \circledast \boldsymbol{F}||^2_F 
+\varphi(\boldsymbol{F}) \right\}, \notag\\
&\hspace{5.4cm}\text{s.t.~} |\mathscr{F}(\mathbfcal{H}_e)|=\tilde{\textbf{H}},
\label{eq:bd}
\end{align}
\noindent where $\tilde{\textbf{H}}$ is the magnitude of the 2D Fourier transform ($\mathscr{F}$) of the PSF and $\varphi$ is a regularization function. Note that $\tilde{\textbf{H}}$ can be assumed known given its straightforward estimation from $\boldsymbol{G}$ by homomorphic \ab{filtering} \cite{Oppenheim75,Taxt1995}. Furthermore, in what follows, $\varphi$ is chosen as the Huber function, due to its ability to characterise complex echogenicity patterns using only one stochastic model \cite{Michailovich2019}: 
\begin{equation}
        \varphi(\boldsymbol{F})= \ \gamma \sum\limits_{n=0}^{N-1}\sum\limits_{m=0}^{M-1}
        \begin{cases}
            |{F}_{n,m}|^2, & {F}_{n,m}\leq a \\
            2a|{F}_{n,m}| - a^2, & \text{otherwise}
        \end{cases},
 \end{equation}
\noindent where $\gamma>0$ is a regularization parameter, and $a>0$ is a parameter balancing the prior between smoothness and sparseness. Moreover, \eqref{eq:bd} can be solved by alternating minimization over $\boldsymbol{F}$ and $\mathbfcal{H}_e$, as follows:
\begin{align}
\label{eq:bd_eq1}
&\hat{\boldsymbol{F}}^{(k+1)}=\mathop{\arg\min_{{\boldsymbol{F}}}} \left\{\frac{1}{2}||\boldsymbol{G}-\dk{\mathbfcal{H}_e^{(k)} }\circledast \boldsymbol{F}||^2_F 
+\varphi(\boldsymbol{F}) \right\},\hspace{0.2cm}
\end{align}
\begin{align}
\label{eq:bd_eq2}
&\hat{\mathbfcal{H}_e}^{(k+1)}=\mathop{\arg\min_{{\mathbfcal{H}_e}}} \left\{\frac{1}{2}||\boldsymbol{G}-\mathbfcal{H}_e\circledast \boldsymbol{F}^{(k+1)}||^2_F \right\}, \notag\\
 &\text{\hspace{5.2cm} s.t.~} |\mathscr{F}(\mathbfcal{H}_e)|=\tilde{\textbf{H}}. 
\end{align}
It is worth mentioning that under the assumption that $\varphi$ is convex, (\ref{eq:bd_eq1}) admits an efficient solution using proximal algorithm, while (\ref{eq:bd_eq2}) is reformulated in the Fourier domain using Parseval's theorem leading to the optimal phase estimation of an all-pass filter that can be efficiently solved by a filter design procedure (\ab{for more details the reader may refer to} \cite{Pei1994, Michailovich2019}).  	    


\section{Proposed BD-RPCA method} \label{sec:BDRPCA}
The proposed algorithm aims at estimating a high resolution blood flow $\boldsymbol{X}$ together with the tissue component $\boldsymbol{T}$ and the PSF $\mathbfcal{H}_e$ from ultrafast US Doppler signals. This novel algorithm is based on a suitable combination of DRPCA and BD methods. The resulting optimization problem to be solved is formulated as follows:
\begin{align} 
\label{eq:bdrpca} 
    [\hat{\boldsymbol{X}},\hat{\mathbfcal{H}_e},\hat{\boldsymbol{T}}] &= \mathop{\arg\min_{{\boldsymbol{X}},{\mathbfcal{H}_e},{\boldsymbol{T}}}}  \left\{ ||\boldsymbol{S}-\mathbfcal{H}_e\circledast \boldsymbol{X}-\boldsymbol{T}||^2_F  \right. \notag\\
      &\hspace{0.3cm} + \left.  \lambda||\boldsymbol{X}||_1 + \rho||\boldsymbol{T}||_*  \right\}, \text{s.t.~~} |\mathscr{F}(\mathbfcal{H}_e)| = \tilde{\textbf{H}}. 
\end{align}
\noindent To solve \eqref{eq:bdrpca}, we propose a two-step alternating algorithm as follows:

\begin{enumerate}[label=\roman*)]
\item For a fixed $\mathbfcal{H}_e$, \eqref{eq:bdrpca} becomes:
\begin{align*} 
    [\hat{\boldsymbol{X}}^{(k+1)},\hat{\boldsymbol{T}}^{(k+1)}]  &= \mathop{\arg\min_{{\boldsymbol{X}},{\boldsymbol{T}}}}  \left\{ ||\boldsymbol{S}-\mathbfcal{H}_e^{(k)} \circledast \boldsymbol{X}-\boldsymbol{T}||^2_F  \right. \notag \\
      &\hspace{2.3cm} + \left.  \lambda||\boldsymbol{X}||_1 + \rho||\boldsymbol{T}||_*  \right\}. 
\end{align*}
This subproblem is solved by RPCA as shown in Section \ref{subsec:RPCA}, \ab{resulting in estimates} $\hat{\boldsymbol{X}}$ and $\hat{\boldsymbol{T}}$.

\item  For fixed $\boldsymbol{X}$ and $\boldsymbol{T}$, and assuming that $\mathbfcal{H}_e$ is spatio-temporally invariant, \ab{the proposed algorithm estimates the PSF} by taking the temporal mean of all clutter filtered frames computed \ab{in} step i), that results into a 2D image. With this in mind, \eqref{eq:bdrpca} is reformulated based on the distributive property of the convolution as follows:	
\begin{align}
&[\hat{\mathbfcal{H}_e}^{(k+1)}] =
\mathop{\arg\min_{{\mathbfcal{H}_e}}} \left\{ ||\overline{\sum\limits_{N_t}}\left(\boldsymbol{S}-\boldsymbol{T}^{(k+1)}\right) \right. \notag \\
      &\hspace{0.5cm}- \left. \mathbfcal{H}_e\circledast \overline{\sum\limits_{N_t}}(\boldsymbol{X}^{(k+1)})||^2_F \right\}, \text{s.t.} |\mathscr{F}(\mathbfcal{H}_e)| = \tilde{\textbf{H}},
\label{eq:bdrpca2}
\end{align} 
\noindent where $\overline{\sum\limits_{N_t}}(\boldsymbol{Z})$ denotes the temporal mean of a 2D matrix $ \boldsymbol{Z}$. This procedure is performed by reshaping the 2D matrix \ab{$\boldsymbol{Z} \in \mathbb{C}^{N_z N_x\times N_t}$} into its corresponding 3D matrix in $\mathbb{C}^{N_z \times N_x\times N_t}$ and then taking the mean along the third dimension. Then, (\ref{eq:bdrpca2}) can be solved by the BD algorithm in Section \ref{subsec:BD} while ignoring the estimation of $\overline{\sum\limits_{N_t}}(\boldsymbol{X}^{(k+1)})$.  
\end{enumerate} 

It should be also noted that RPCA is used to initialize the values of the blood and tissue, prior to the process of the proposed algorithm, allowing an efficient convergence speed-up. The pseudo algorithm \dk{related to} the proposed blood flow retrieval method, named BD-RPCA, is given in Algorithm \ref{alg:drpca}. 

\begin{algorithm}
 \label{alg:drpca}
\SetAlgoLined
\KwIn{observed Casorati matrix $\boldsymbol{S}$.}
\textbf{Initialize:} $tol=10^{-6}$, $[\boldsymbol{X}^{(0)},\boldsymbol{T}^{(0)}]$ = \text{RPCA}($\boldsymbol{S}$); \\
 \While{$||\boldsymbol{X}^{(k+1)}-\boldsymbol{X}^{(k)}||_F > tol$}{   
   \begin{enumerate}
    \item compute temporal mean: $\boldsymbol{M}_{\boldsymbol{ST}}^{(k+1)}=\overline{\sum\limits_{N_t}}\left(\boldsymbol{S}-\boldsymbol{T}^{(k)}\right)$ 
   	\item estimate PSF: $[\mathbfcal{H}_{e}^{(k+1)}]$ = \text{BD}$\left(\boldsymbol{M}_{\boldsymbol{ST}}^{(k+1)}\right)$
    \item update: $[\boldsymbol{X}^{(k+1)},\boldsymbol{T}^{(k+1)}]$ = \text{DRPCA}$\left(\boldsymbol{S},\mathbfcal{H}_{e}^{(k+1)}\right)$     
     \end{enumerate}           
 }     
 \KwOut{high-resolution blood $\boldsymbol{X}^{(k+1)}$ and estimated PSF $\mathbfcal{H}_{e}^{(k+1)}$.}
 \caption{BD-RPCA}
\end{algorithm}

\section{Numerical Results} \label{sec:Results}
This section presents several numerical experiments on both simulated and \textit{in vivo} US data to illustrate the contribution of BD-RPCA over the three existing methods SVD \cite{demene_spatiotemporal_2015}, RPCA and DRPCA \cite{Shen2019}. All the experiments were conducted using MATLAB R2019b on a computer with Intel(R) Core(TM) i5-8500 CPU @3.00 GHz and 16GB RAM. 

\begin{figure}[!htb]
\begin{minipage}[b]{\linewidth}
\centering
\centerline{\includegraphics[width=0.75\linewidth, height = 6cm]{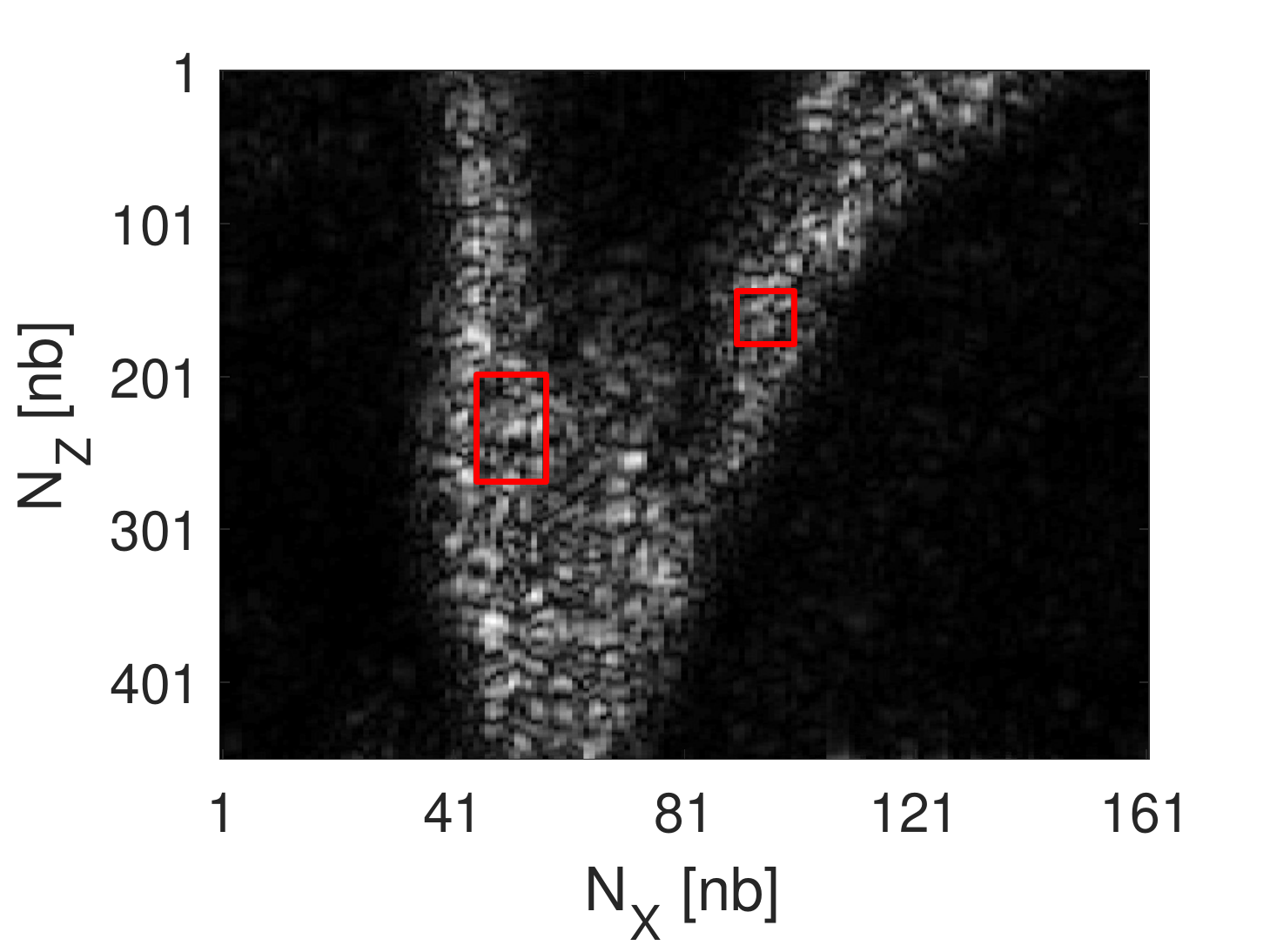}}
\end{minipage}
\caption{Simulated B-mode image. Note that $N_z[nb]$ and $N_x[nb]$ denote the lateral and axial distances in number of samples.}
\label{fig:simulation_Bmode}
\end{figure} 

\subsection{Simulation results}
In this simulation, we consider a static vessel whose dimensions $N_z \times N_x \times N_t$ are respectively $451\times 161 \times 400$ samples, as shown in Fig.~\ref{fig:simulation_Bmode}. Such a vessel is associated to the static tissue generated from randomly distributed scatterers with Gaussian random amplitudes. The US frame rate was set to $12.8$ kHz. The pixel increments in the directions $x$ and $z$ are respectively $d_x = 0.0333$ cm and $d_z = 0.0086$ cm while the sampling frequency was set to $9$ MHz. In addition, two moving rectangles of sizes $12\times70$ and $10\times35$ pixels were simulated inside the vessel so that their interior points randomly move \dk{using \textit{circshift} MATLAB function in order to mimic the blood flow}. The resulting image was convolved with an experimentally measured PSF to simulate the loss of spatial resolution. \ab{Note that this measured PSF was also used within DRPCA in the estimation process, as proposed in \cite{Shen2019}}.

\begin{table}[!htb]
\centering
\renewcommand{\arraystretch}{1.3}
\caption{Optimal setting of $\lambda$ and $\mu$ for the simulation}
\label{tab:lambd_mu}
\begin{tabular}{c|c|c|c|}
\cline{2-4}
                       & \bfseries RPCA & \bfseries DRPCA  &  \bfseries BD-RPCA \\ \hline
\multicolumn{1}{|l|}{$\lambda$}  &  0.0111 & 0.0111 & 0.0037 \\ \hline
\multicolumn{1}{|l|}{$\mu$ } &  0.1113 & 0.0223 &   0.0074 \\ \hline
\end{tabular}
\end{table} 

\subsubsection{Hyperparameter tuning}

To ensure a fair comparison, hyperparameters associated with each method were tuned by cross-validation to their best possible values. For SVD, the clutter and blood rank thresholds as introduced in (\ref{eq:B_SVD}) were chosen as $T_c = 2$ and $T_b = 15$, respectively. For the RPCA-based methods, we recall that $\lambda$ and $\rho$ reflect the compromise between the blood sparsity and the tissue low-rankness while $\mu$ handles the algorithm convergence rate. As suggested in \cite{Wright_robust_2009} for the general RPCA problem,  $\rho$ was set to $1$ while the reference values $\lambda_{\text{ref}}=\frac{1}{\sqrt{\max(Nz \times Nx,Nt)}}$ and $\mu_{\text{ref}}=\mu_0 \times \lambda_{\text{ref}}$, where $\mu_0$ is a fixed multiplier: $10$ for RPCA and $2$ for DRPCA or BD-RPCA, were used in order for the optimal tuning to be carried out more efficiently. Given this consideration, the best $\lambda$ and $\mu$ values selected for this simulation are shown in Table \ref{tab:lambd_mu}. It is interesting to note that the $\lambda$ associated within RPCA or DRPCA is about $3$ times \ab{higher} than the one used within BD-RPCA. Moreover, when using BD, $\gamma$ was set to $0.002$ and $a$ to $0.05$. 

\ab{Note also that} the most common Power Doppler image for representing the retrieved blood flow was used \ab{for visualisation purpose}. The Power Doppler image measured in dB, \dk{denoted by $\boldsymbol{I}_{PD}$}, is calculated from the estimated blood flow $\boldsymbol{B}$, for a given position $(x,z)$, as follows:
\begin{equation*}
\boldsymbol{I}_{PD}(x,z) = 10\log10\left(\frac{1}{N_t}\sum\limits_{k=1}^{N_t}\boldsymbol{B}(x,z,k)^{2}\right).
\end{equation*}  

\begin{figure*}[!]
\begin{minipage}[b]{.185\linewidth}
\centering
\centerline{\includegraphics[width=\linewidth,  height = 5.5cm]{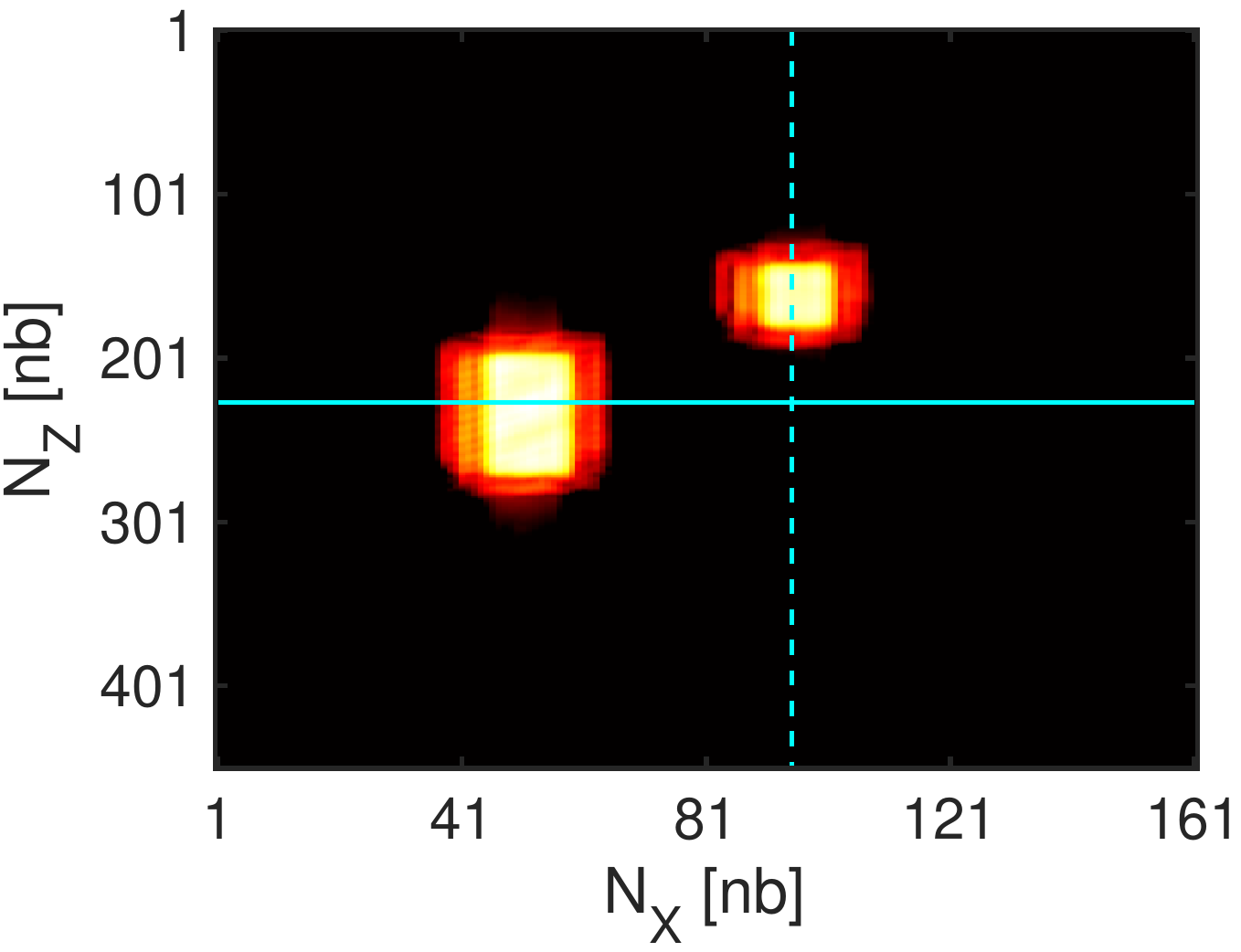}}
\centerline{{\scriptsize(a) }} 
\end{minipage}
\begin{minipage}[b]{0.185\linewidth}
\centering
\centerline{\includegraphics[width=\linewidth,  height = 5.5cm]{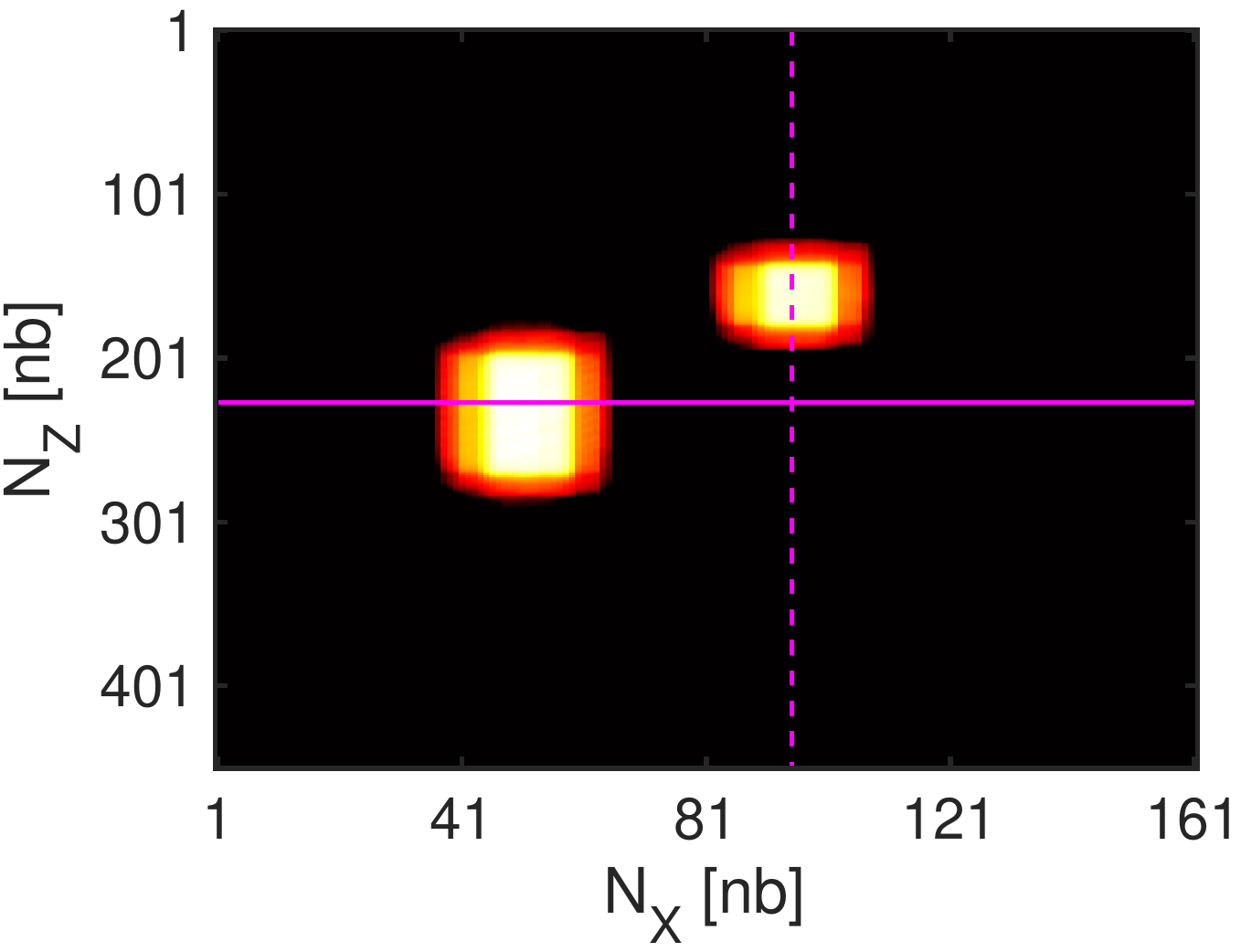}}
\centerline{{\scriptsize(b) }} 
\end{minipage}
\begin{minipage}[b]{0.185\linewidth}
\centering\centerline{\includegraphics[width=\linewidth,  height = 5.5cm]{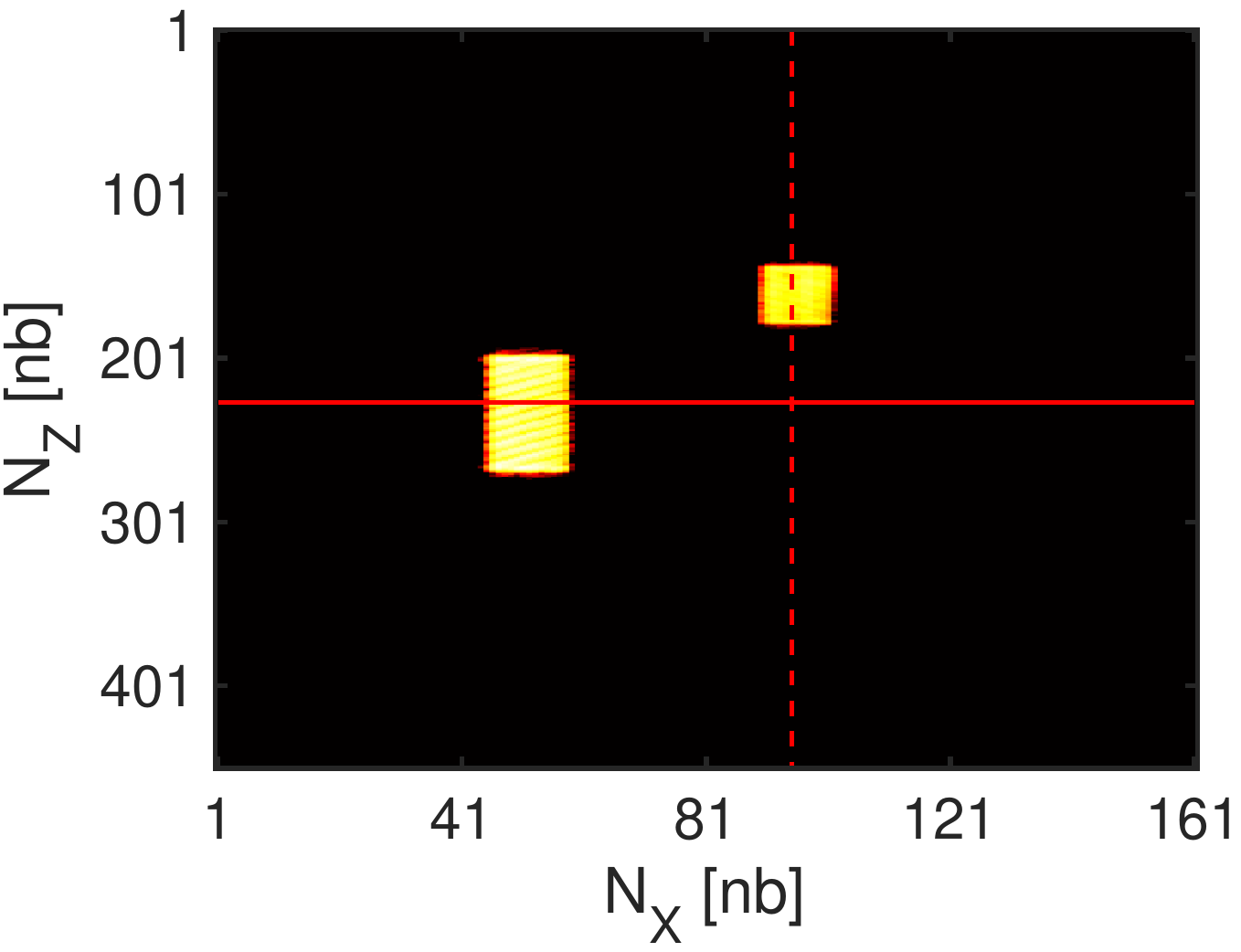}}
\centerline{{\scriptsize (c) }} 
\end{minipage}
\begin{minipage}[b]{0.185\linewidth}
\centering
\centerline{\includegraphics[width=\linewidth,  height = 5.5cm]{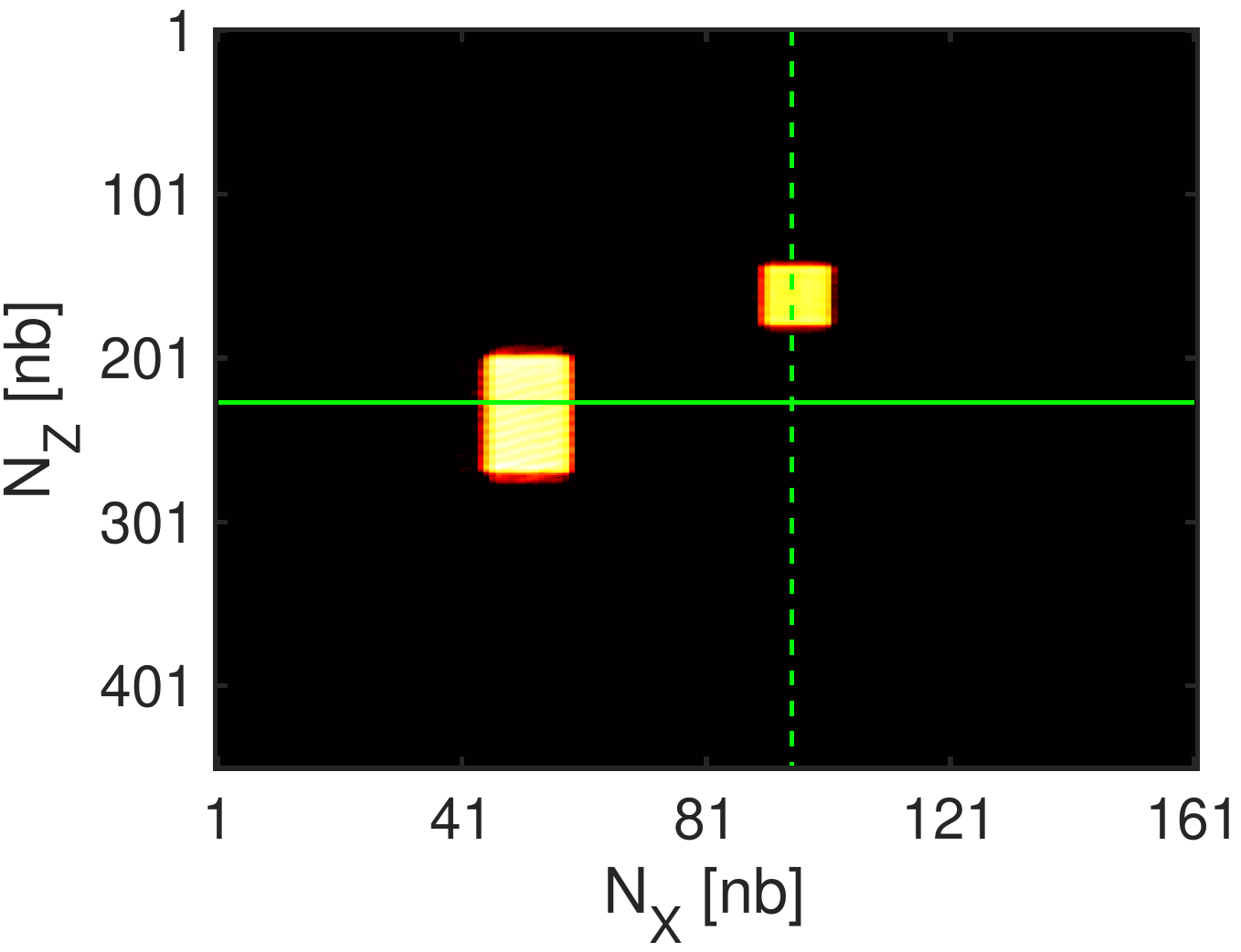}}
\centerline{{\scriptsize (d) }} 
\end{minipage}
\begin{minipage}[b]{.185\linewidth}
\centering
\centerline{\includegraphics[width=\linewidth,  height = 5.5cm]{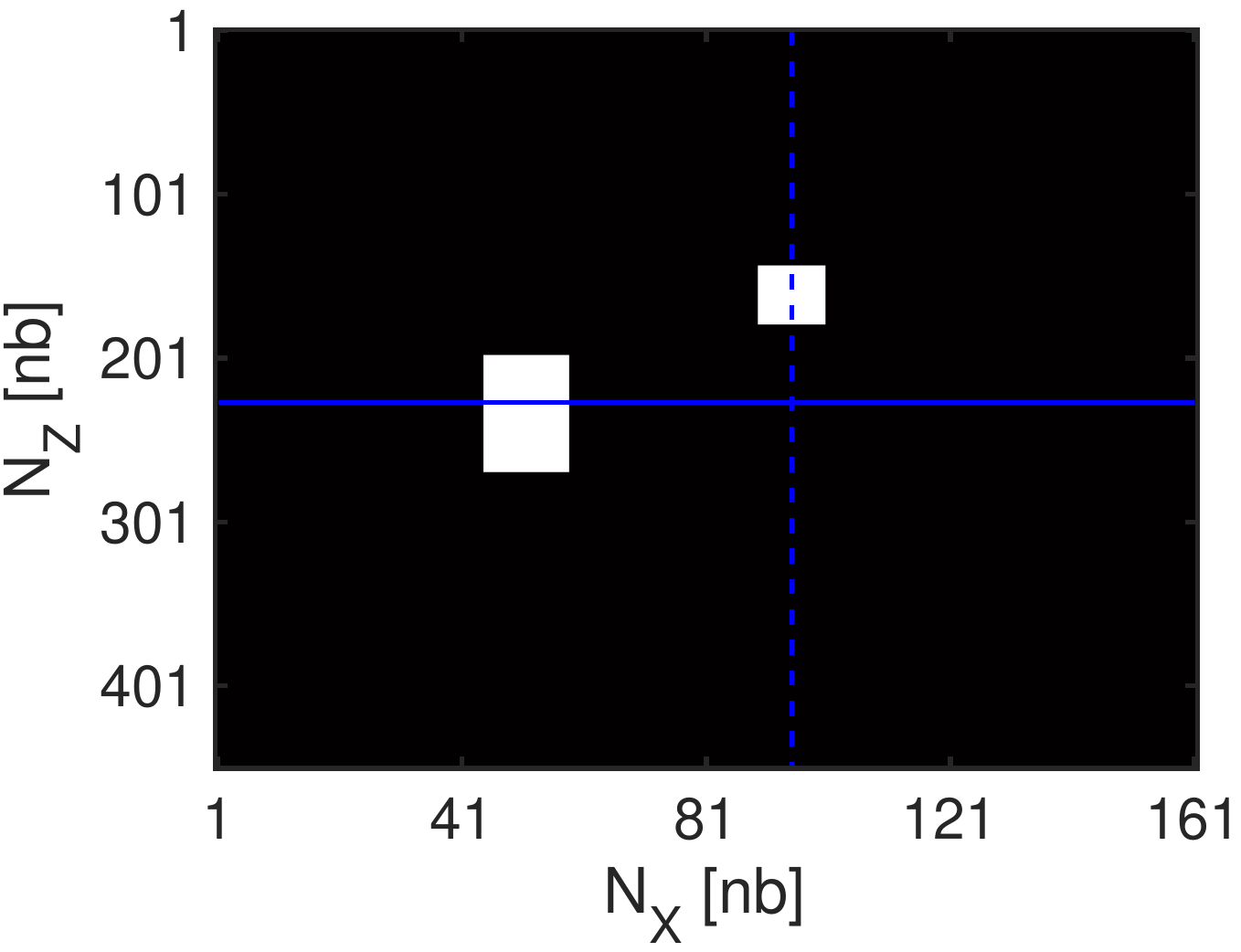}}
\centerline{{\scriptsize (e) }} 
\end{minipage}
\begin{minipage}[b]{.035\linewidth}
\centering
\centerline{\includegraphics[trim={12cm 0cm 0 0},clip, width=\linewidth,  height = 5.5cm]{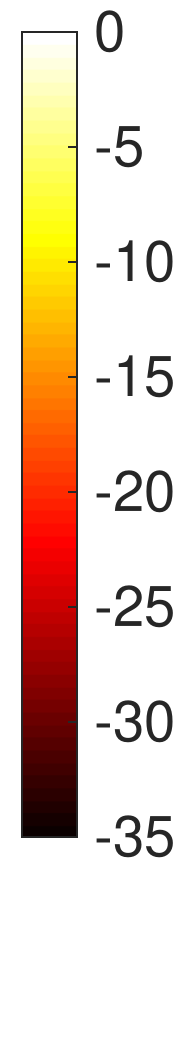}}
\centerline{} 
\end{minipage}
\caption{Power Doppler images \ab{estimated from the simulated data in the noiseless case} with: (a) SVD; (b) \ab{RPCA}; (c) DRPCA; (d) BD-RPCA and (e) the corresponding ground truth. All the images are displayed with a dynamic range of $35$ dB.}
\label{fig:simulation}
\end{figure*}

\begin{figure*}[!htb]
\begin{minipage}[b]{0.48\linewidth}
\centering
\centerline{\includegraphics[width=0.9\textwidth, height = 6cm]{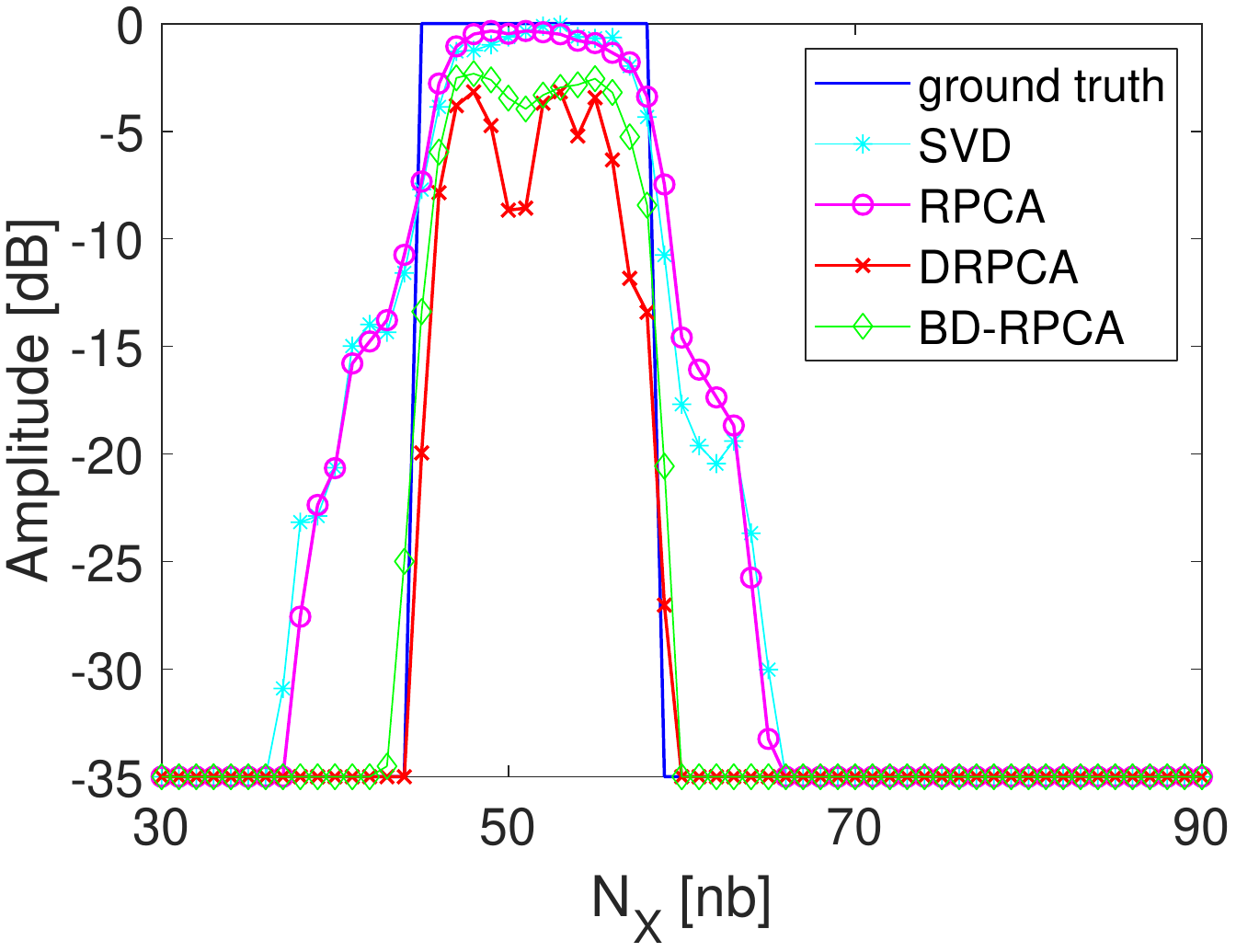}}
\centerline{{\scriptsize(a)}} 
\end{minipage}
\begin{minipage}[b]{0.48\linewidth}
\centering
\centerline{\includegraphics[width=0.9\textwidth, height = 6cm]{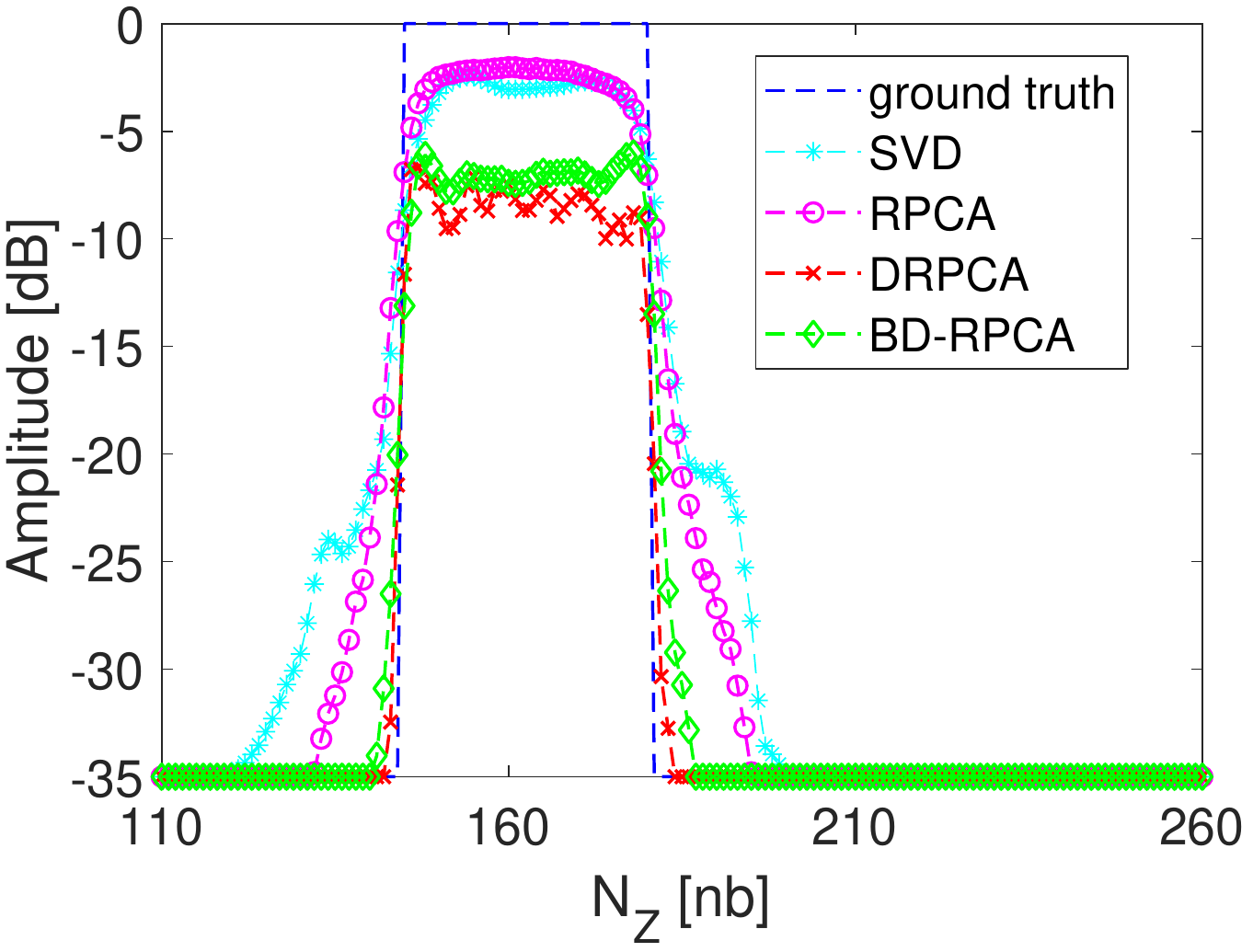}}
\centerline{{\scriptsize(b)}} 
\end{minipage}
\caption{(a) Lateral cross-profiles at the \emph{solid horizontal} lines of Fig. \ref{fig:simulation}; (b) Axial cross-profiles at the \emph{dashed vertical} lines of Fig. \ref{fig:simulation}.}
\label{fig:cut_off}
\end{figure*}

\subsubsection{Noise-free case}
Let us first display, in Fig.~\ref{fig:simulation}, the estimated Power Doppler results given by the four studied methods and the ground truth \ab{for the noiseless case}. From these results, one may remark that Power Doppler images estimated by both SVD and RPCA are quite noisy and \ab{blurred}, whereas the one obtained by BD-RPCA presents considerably sharper edges around the two rectangles, very similar to DRPCA. This observation is further confirmed by the plots in Fig. \ref{fig:cut_off}, showing respectively the lateral and axial variations at two different positions as depicted in Fig. \ref{fig:simulation}. From these plots, it can be seen that the lateral and axial profiles given by BD-DRPCA have sharp slopes extremely close to the ones given by DRPCA as well as to the ground truth, whereas those obtained by SVD and RPCA are obviously much \ab{less resolved}. 

\begin{table}[!htb]
\centering
\renewcommand{\arraystretch}{1.3}
\caption{Quality assessment for the noiseless case}
\label{tab:mean_metrics}
\begin{tabular}{c|c|c|c|c|}
\cline{2-5}
                            & \bfseries SVD & \bfseries RPCA & \bfseries DRPCA & \bfseries BD-RPCA \\ \hline
\multicolumn{1}{|c|}{NRMSE} &  0.0890 & 0.0832 & 0.0409 & 0.0411  \\ \hline
\multicolumn{1}{|c|}{PSNR [dB]}  &  21.092 & 21.685 & 27.840 & 27.800  \\ \hline
\end{tabular}
\end{table}

Additionally, since the ground truth of simulated images is available, the blood flow retrieval performance was quantitatively evaluated in terms of normalized root mean square error (NRMSE) \dk{and} peak signal-to-noise ratio (PSNR). \ab{These} metrics are defined as follows:
\begin{align*}
&\text{NRMSE} =  \sqrt{\frac{||\boldsymbol{I}_{PD}-\boldsymbol{\hat{I}}_{PD}||^2_F}{||\boldsymbol{I}_{PD}||^2_F}},\\
&\text{PSNR}_{[\text{dB}]}  = 10\log10 \left(\frac{{d}_{\max }^2}{\text{MSE}}\right),
\end{align*}
where $\boldsymbol{\hat{I}}_{PD}$ and $\boldsymbol{I}_{PD}$ are respectively the estimated Power Doppler image and its corresponding ground truth. Regarding the PSNR formula, $\text{MSE} = \frac{1}{N_zN_x}||\boldsymbol{I}_{PD}-\boldsymbol{\hat{I}}_{PD}||^2_F$ is the mean square error (MSE) between the two images while ${d}_{\max } = 35$ denotes the maximum pixel value, i.e., the dynamic range, of the image. The quantitative results reported in Table \ref{tab:mean_metrics} clearly exhibit a very slightly lower estimation accuracy of BD-RPCA \ab{compared to} \ab{DRPCA} and considerably better than the two other \ab{approaches}, confirming the coherence with the previous visual observation. \ab{Note however the noticeable difference between BD-RPCA and DRPCA: whilst the proposed approach is unsupervised and able to estimate the PSF jointly with the blood flow, the latter uses the true PSF used to simulate the Doppler data.}

\begin{figure}[!htb]
\begin{minipage}[b]{0.49\linewidth}
\centering
\centerline{\includegraphics[width=\textwidth, height = 5cm]{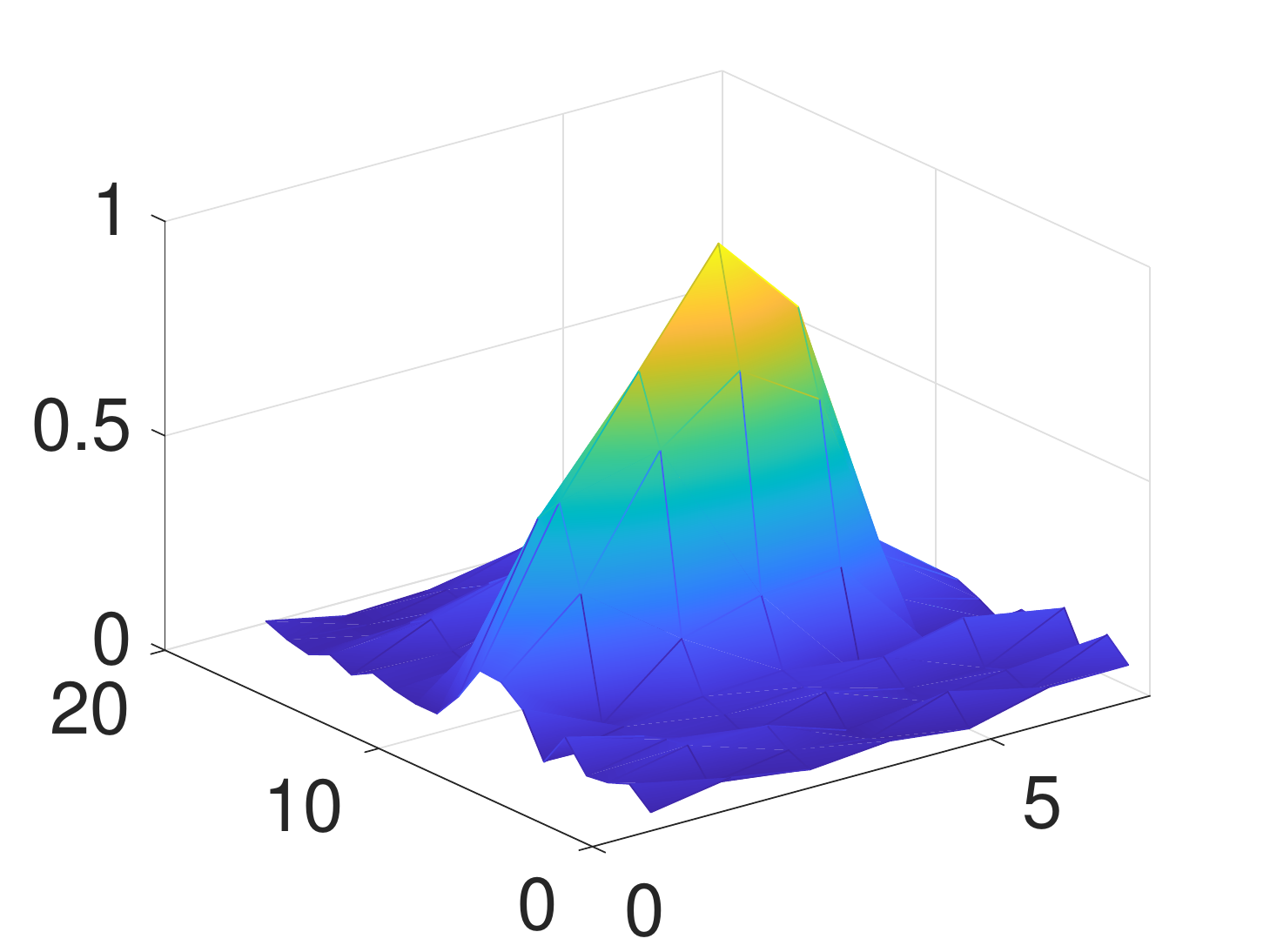}}
\centerline{{\scriptsize(a)}} 
\end{minipage}
\begin{minipage}[b]{0.49\linewidth}
\centering
\centerline{\includegraphics[width=\textwidth, height = 5cm]{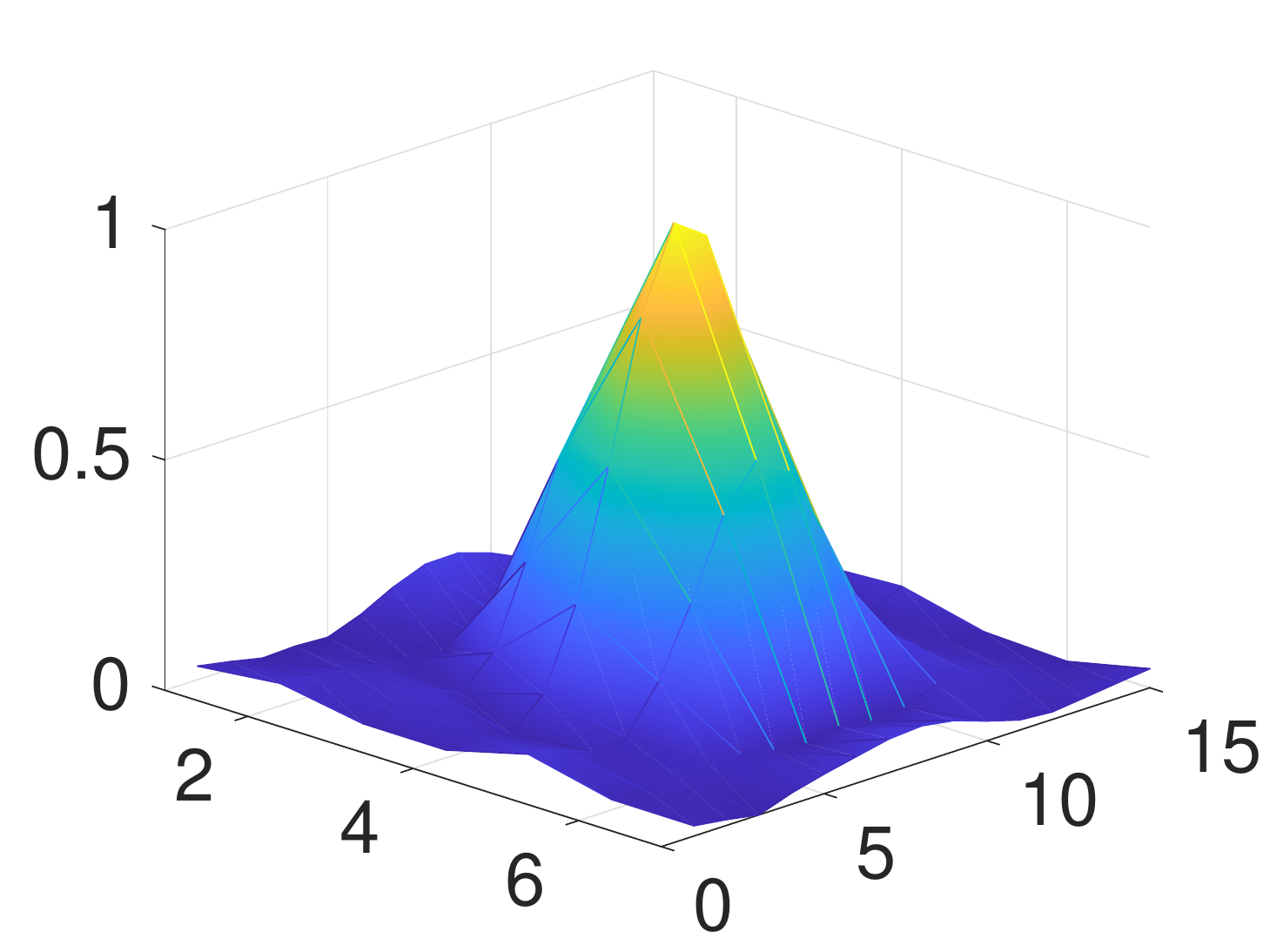}}
\centerline{{\scriptsize(b)}} 
\end{minipage}
\caption{(a) PSF $\mathbfcal{H}_m$ used to simulate the Doppler data; (b) estimated PSF $\mathbfcal{H}_e$ by BD-RPCA. Note that these two PSFs are normalized between $0$ and $1$ for visualization comparison purpose.}
\label{fig:PSFs}
\end{figure}

\ab{Finally}, we depict in Fig. \ref{fig:PSFs} the estimated PSF obtained with the proposed algorithm, in comparison with the experimentally measured PSF \ab{used to simulate the data}. \ab{One may remark the ability of the the proposed blind deconvolution approach to estimate the unknown PSF with reasonable accuracy, thus proving its interest} for ultrafast US imaging. 

\begin{figure*}[!htb]
\begin{minipage}[b]{0.49\linewidth}
\centering
\centerline{\includegraphics[width=0.9\textwidth, height = 6.5cm]{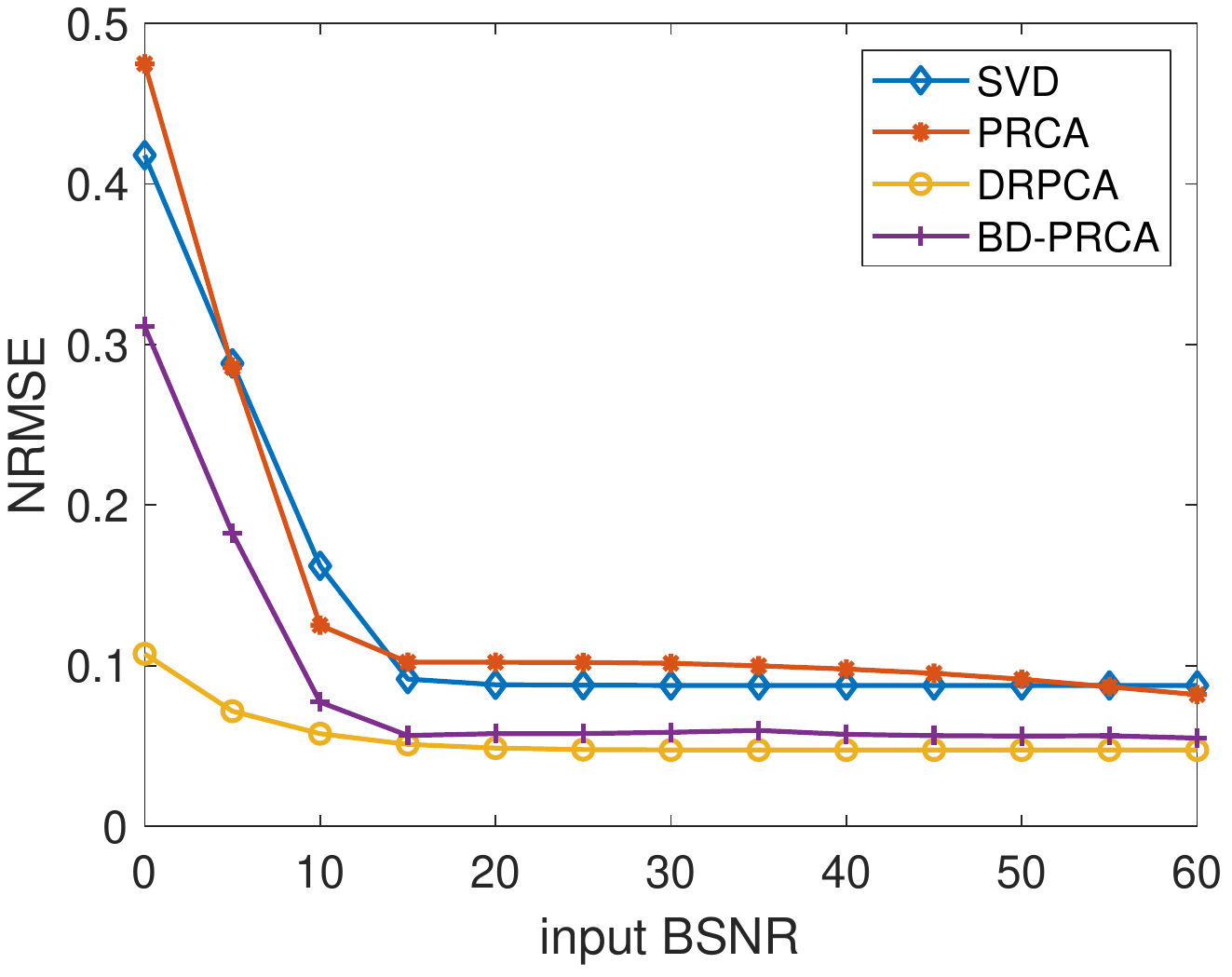}}
\centerline{{\scriptsize(a)}} 
\end{minipage}
\begin{minipage}[b]{0.48\linewidth}
\centering
\centerline{\includegraphics[width=0.9\textwidth, height = 6.5cm]{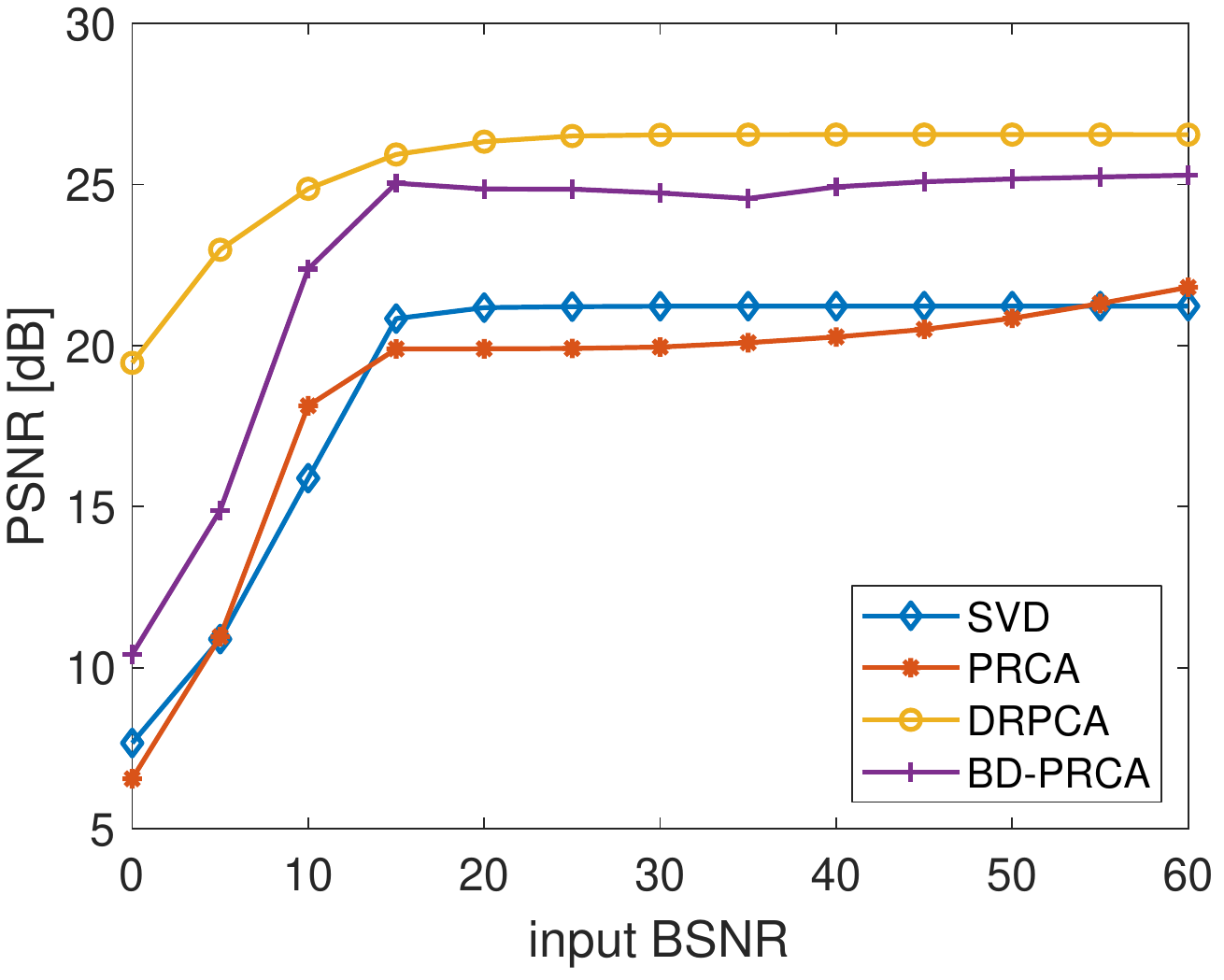}}
\centerline{{\scriptsize(b)}} 
\end{minipage}
\caption{Quantitative measures of different retrieval techniques as a function of input BSNR in the noisy case, in terms of: (a) NRMSE; (b) PSNR.}
\label{fig:metrics_noisy}
\end{figure*}

\subsubsection{Noisy case}
To further test the robustness of the different blood flow recovery methods in the presence of noise, each frame \ab{generated as explained previously was} contaminated by an additive white Gaussian noise (AWGN), as done in  \cite{Zhao2016}. The AWGN level is characterized by the blurred signal-to-noise ratio (BSNR) expressed in dB as follows:
\begin{equation*}
\text{BSNR}_{[\text{dB}]} = 10\log10 \left( \frac{||\boldsymbol{H}\boldsymbol{X}-E(\boldsymbol{H}\boldsymbol{X})||^2_F}{N\sigma^2_n}\right)
\end{equation*}
where $E$ stands for the empirical average, $N$ for the total number of image pixels and $\sigma^2_n$ for the noise variance. Note that the hyperparameter setting associated with each technique was kept the same as for the noiseless case and that the evaluation results represent the averages of the estimations computed by repeating each simulation $100$ times.

\ab{Fig. \ref{fig:metrics_noisy} illustrates the variation of the two evaluation \dk{metrics} NRMSE and PSNR obtained when applying the different retrieval techniques to the simulated data with varying input BSNR values (from $0$ dB to $60$ dB). Examining these plots, it is firstly noticeable that as the input BSNR value increases, the evaluation results given by each reconstruction technique go down in terms of NRMSE and go up in terms of PSNR, and then all stagnate when some particular value for BSNR (about $15$ dB) is reached. It can be easily predicted that each of these curves will get closer and closer to the corresponding upper bound as depicted in Table \ref{tab:mean_metrics} for the noiseless case if the input BSNR value continues to rise. Moreover, one may remark that DRPCA still yields the best blood flow recovery accuracy whatever the input BSNR value due to the use of the true measured PSF. However, for BSNR values larger or equal to $15$ dB, BD-RPCA that relies on estimated PSF exhibits an asymptotic behaviour with a just slightly lower performance. Specifically, the average difference between these two methods from $15$ dB to $60$ dB is of approximately $10^{-2}$ in terms of NRMSE and roughly $1.5$ dB in terms of PSNR. Furthermore, it is obvious from Fig. \ref{fig:metrics_noisy} that BD-RPCA substantially outperforms SVD and RPCA regardless of the noise level.}

To conclude, all the above simulated results confirm the quasi-equivalent ability of BD-RPCA to DRPCA in estimating high-resolution and high-sensitivity blood flow of ultrafast Doppler data with moderate noise levels, i.e., BSNR values of $15-20$ dB and higher, with the advantage of the former that an \textit{a priori} knowledge of the PSF is not required.

\begin{figure*}[!htb]
\begin{minipage}[b]{.23\linewidth}
\centering
\centerline{\includegraphics[width=0.99\textwidth, height = 6.5cm]{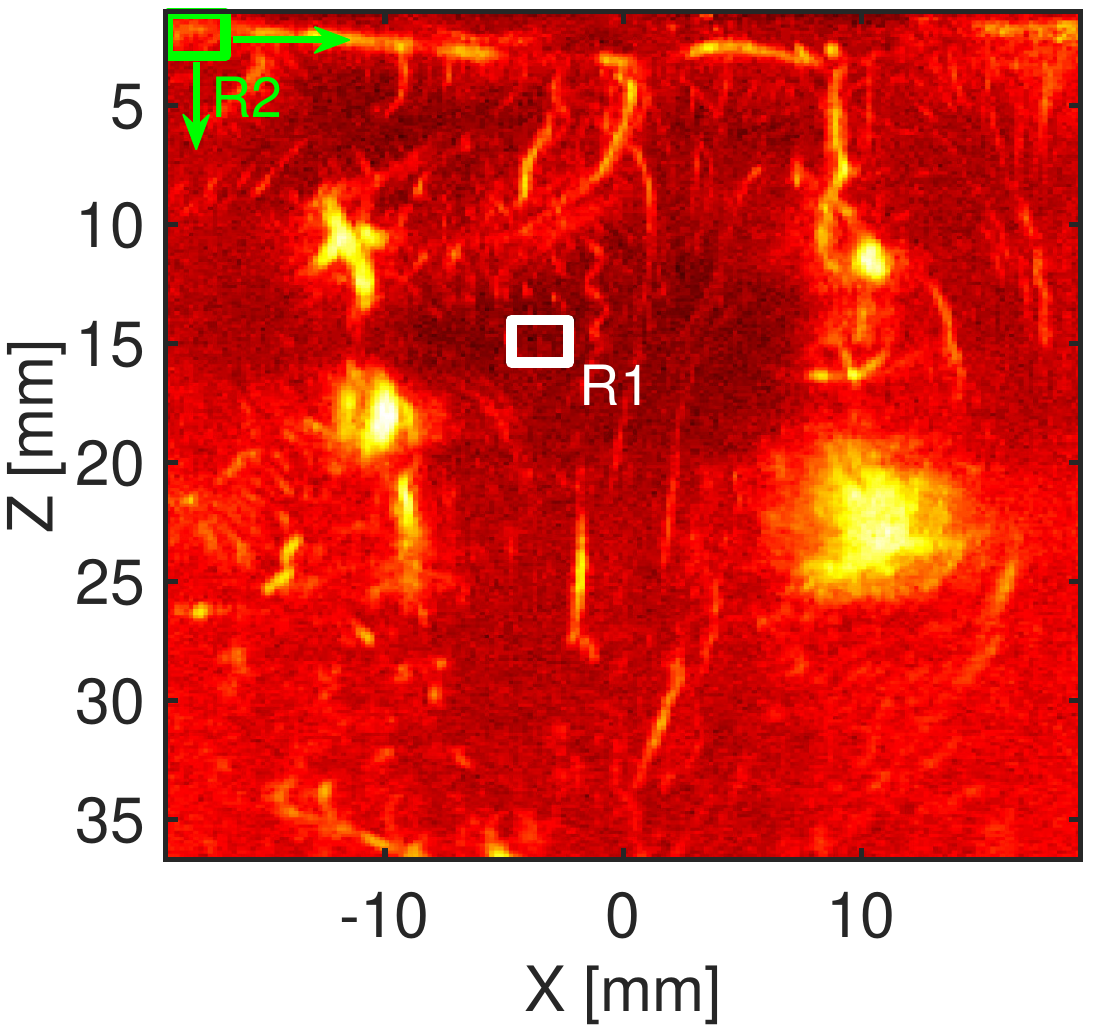}}
\centerline{{\scriptsize(a) $T_c = 100, T_b = 150$}} 
\end{minipage}
\begin{minipage}[b]{.23\linewidth}
\centering
\centerline{\includegraphics[width=0.99\textwidth, height = 6.5cm]{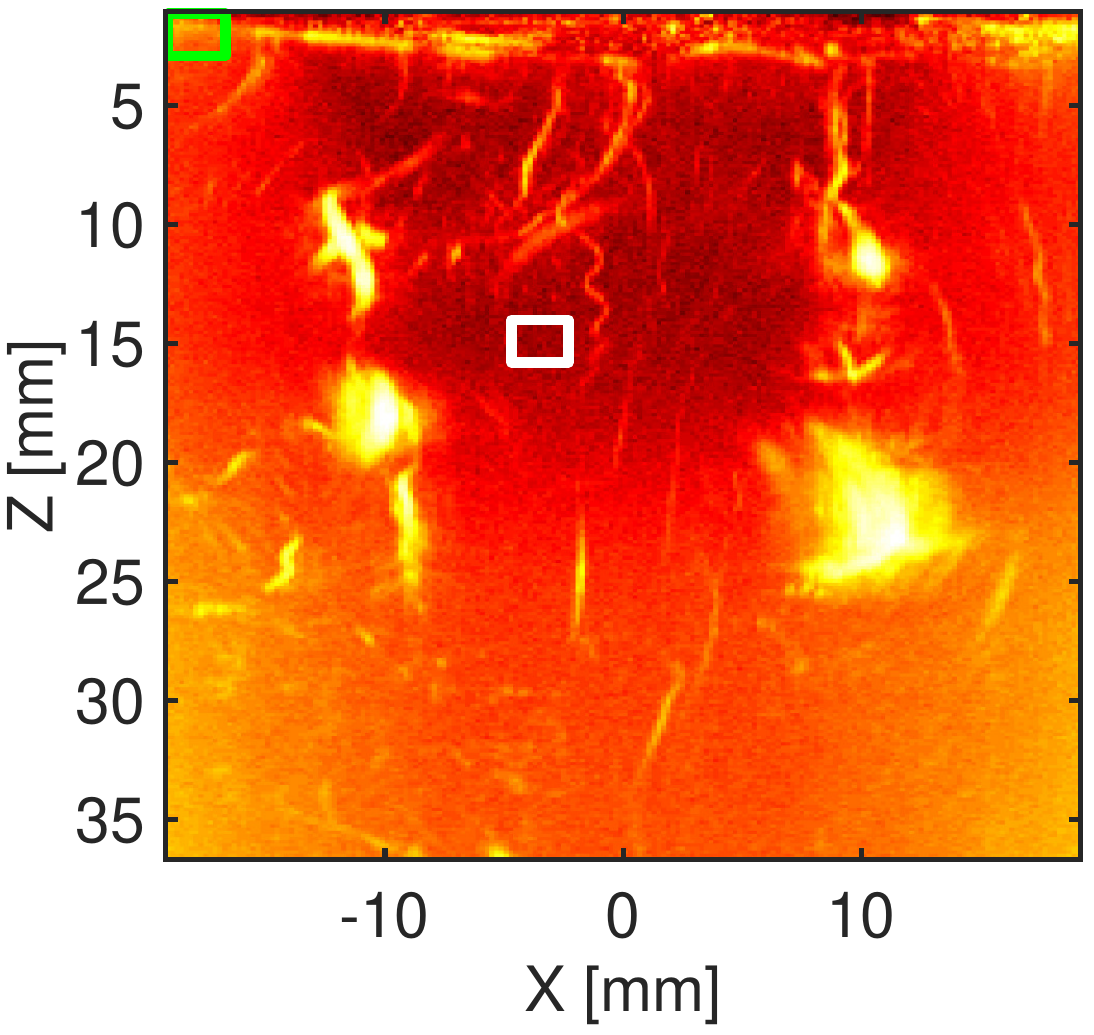}}
\centerline{{\scriptsize(b) $\lambda = 0.0058$, $\mu = 0.0582$}} 
\end{minipage}
\begin{minipage}[b]{0.23\linewidth}
\centering
\centerline{\includegraphics[width=0.99\textwidth, height = 6.5cm]{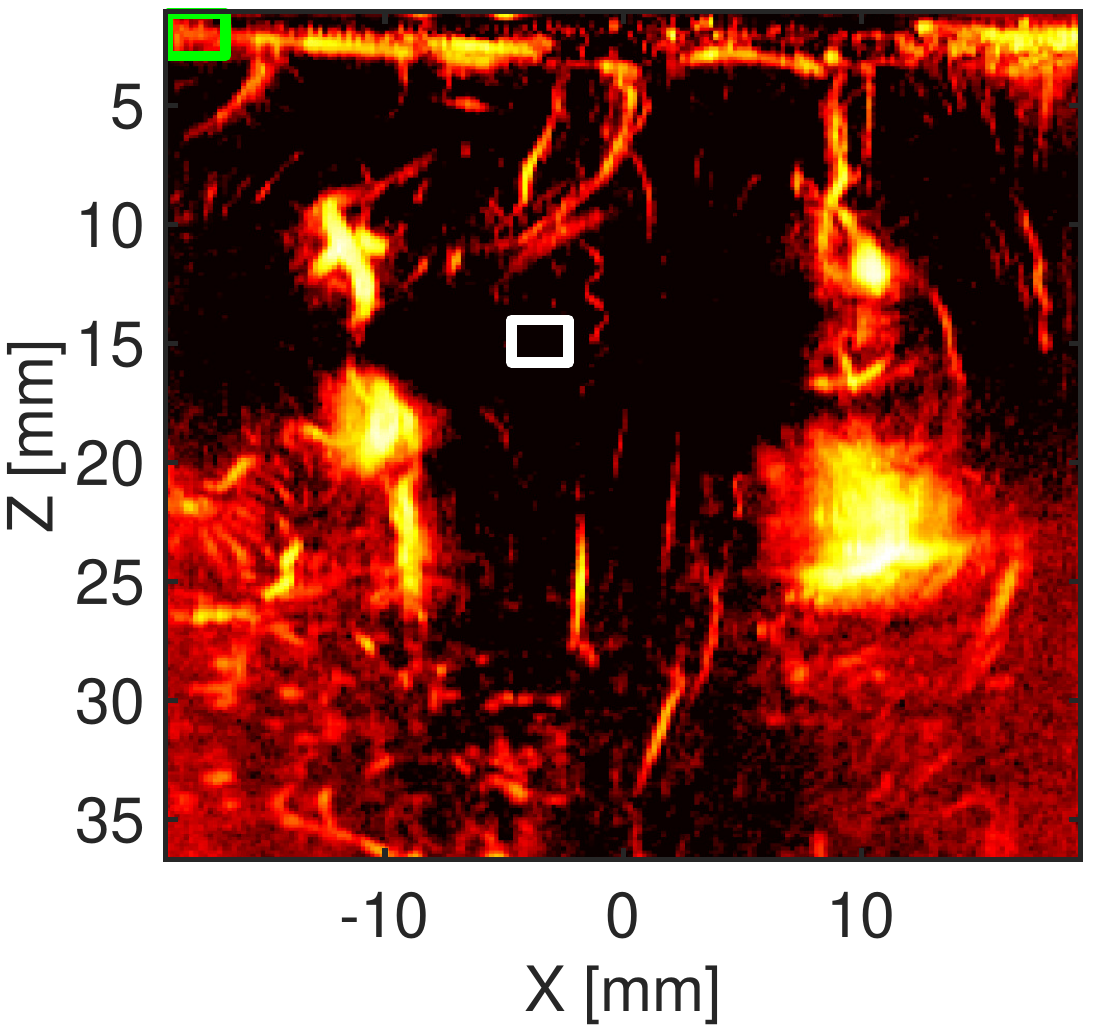}}
\centerline{{\scriptsize(c) $\lambda = 0.0058$, $\mu = 0.0116$}} 
\end{minipage} 
\begin{minipage}[b]{0.23\linewidth}
\centering
\centerline{\includegraphics[width=0.99\textwidth, height = 6.5cm]{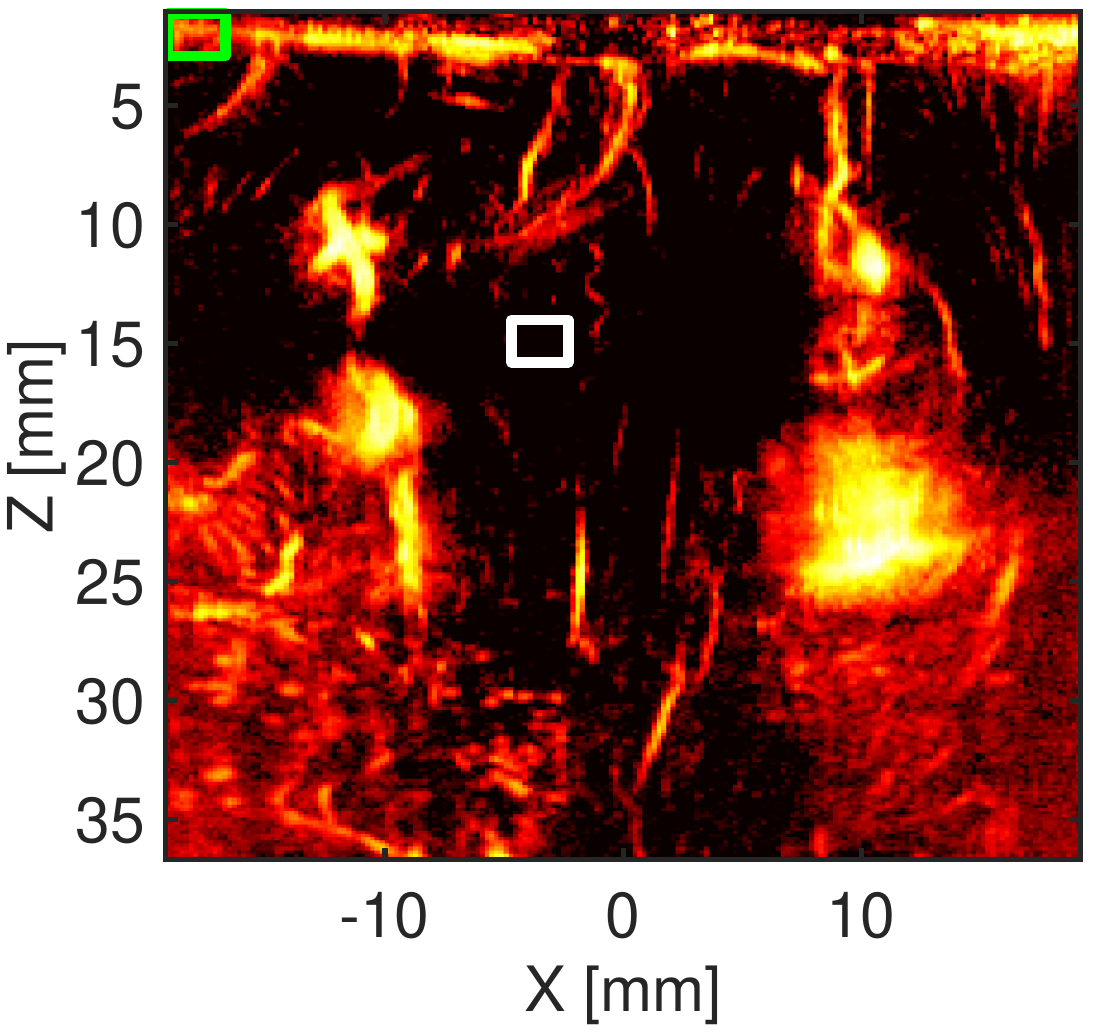}}
\centerline{{\scriptsize (d) $\lambda = 0.0045$, $\mu = 0.0090$}} 
\end{minipage} 
\begin{minipage}[b]{0.05\linewidth}
\centering
\centerline{\includegraphics[trim={11.2cm 0.03cm 0 0},clip, width=0.7\textwidth, height = 6.6cm]{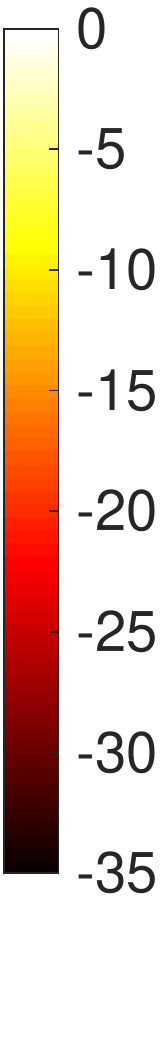}}
\centerline{} 
\end{minipage}
\\
\begin{minipage}[b]{.23\linewidth}
\centering
\centerline{\includegraphics[width=0.99\textwidth, height = 6.5cm]{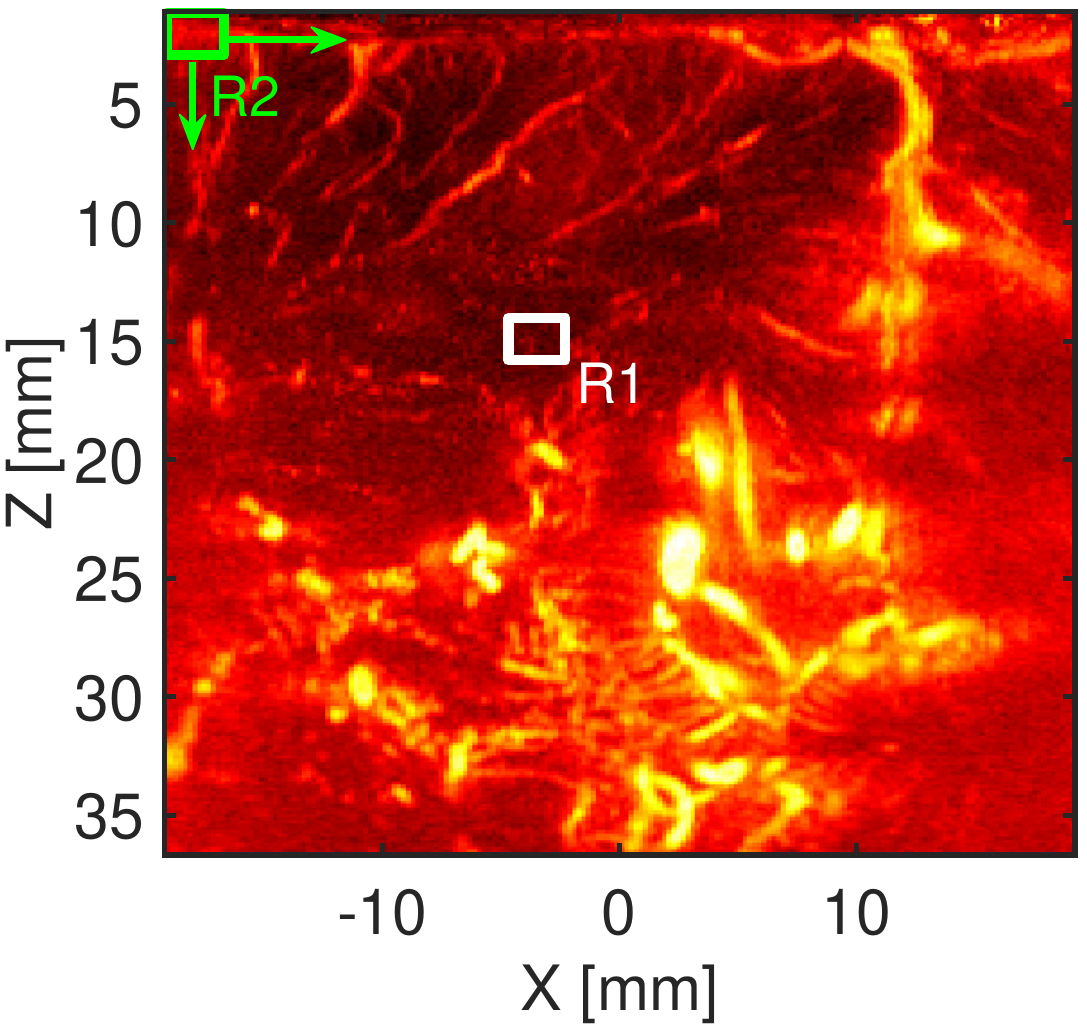}}
\centerline{{\scriptsize(e) $T_c = 100, T_b = 200$}} 
\end{minipage}
\begin{minipage}[b]{.23\linewidth}
\centering
\centerline{\includegraphics[width=0.99\textwidth, height = 6.5cm]{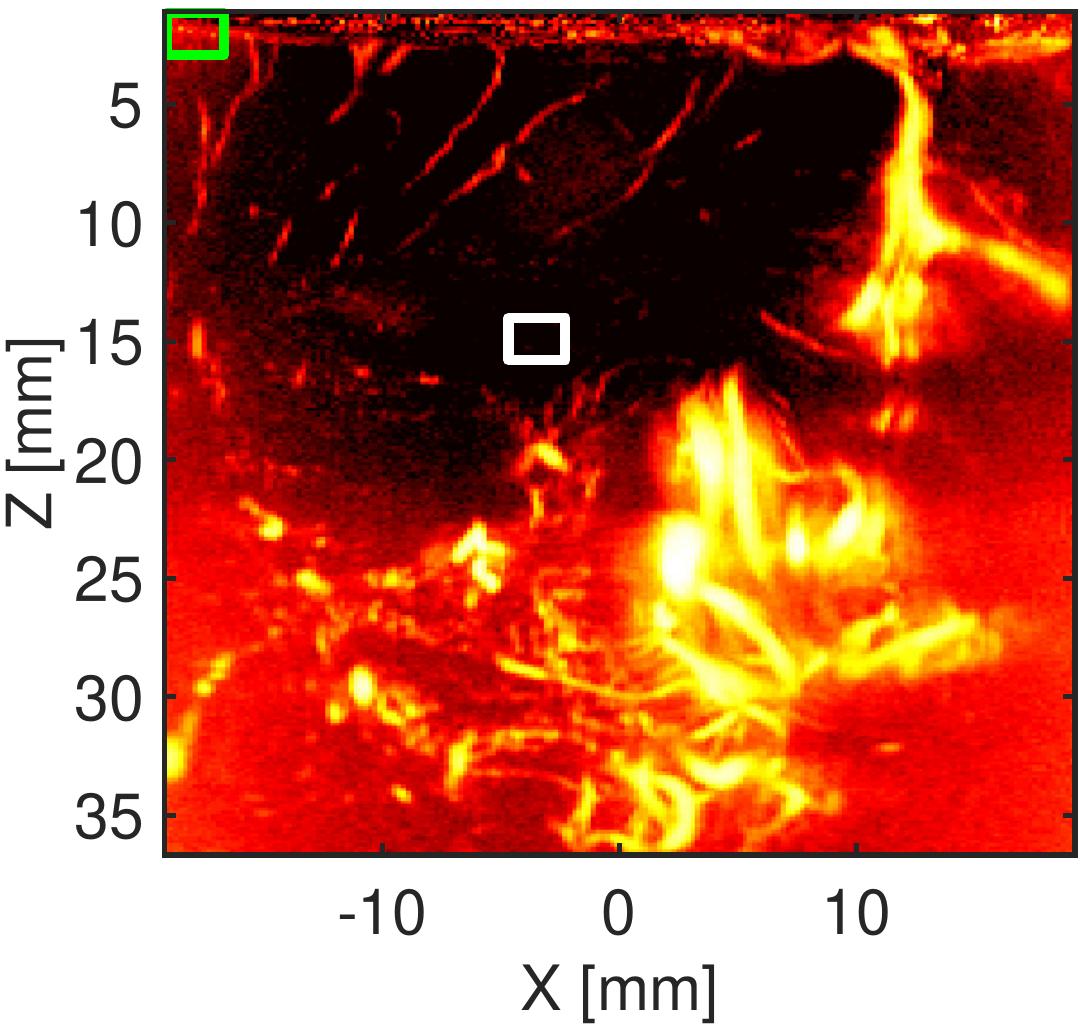}}
\centerline{{\scriptsize(f) $\lambda = 0.0054$, $\mu = 0.0537$}} 
\end{minipage}
\begin{minipage}[b]{0.23\linewidth}
\centering
\centerline{\includegraphics[width=0.99\textwidth, height = 6.5cm]{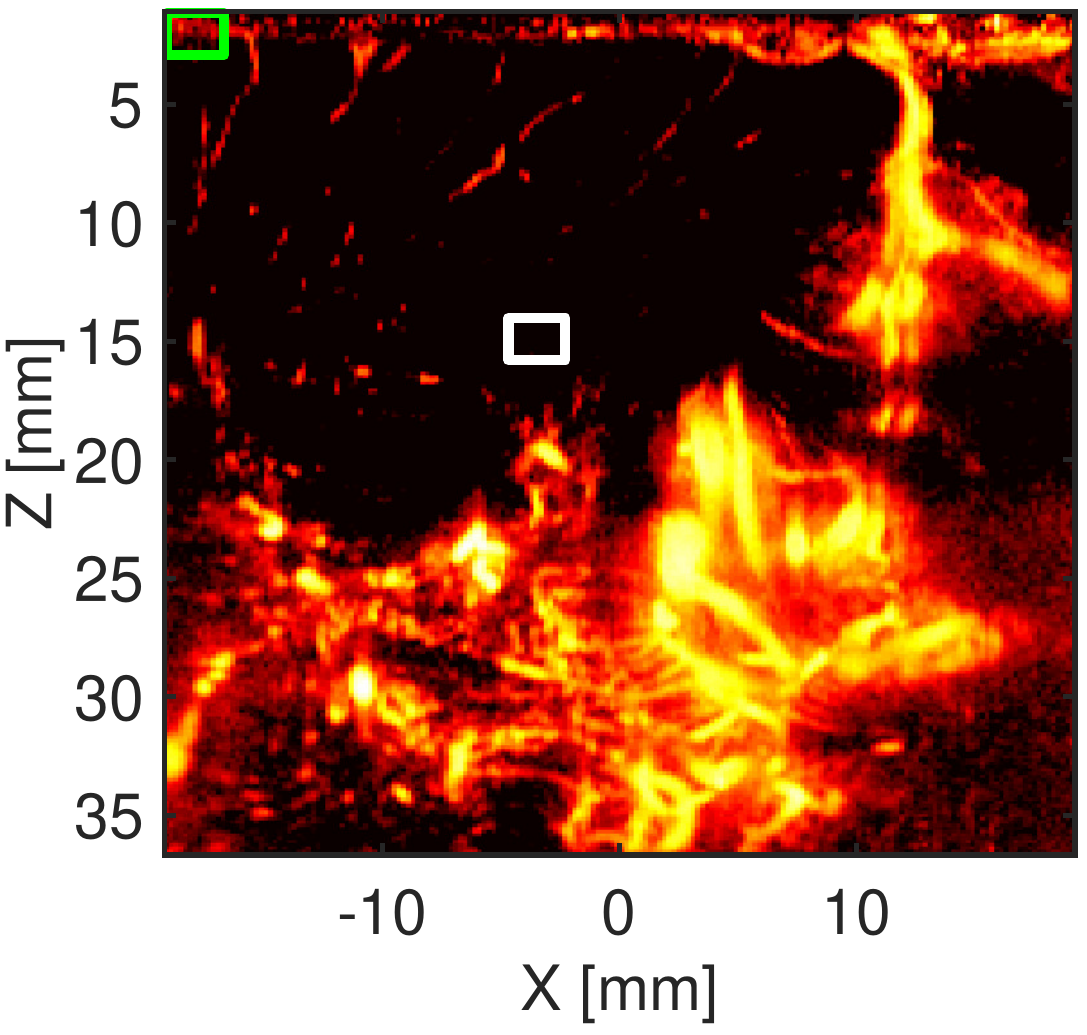}}
\centerline{{\scriptsize(g) $\lambda = 0.0054$, $\mu = 0.0107$}} 
\end{minipage}
\begin{minipage}[b]{0.23\linewidth}
\centering
\centerline{\includegraphics[width=0.99\textwidth, height = 6.5cm]{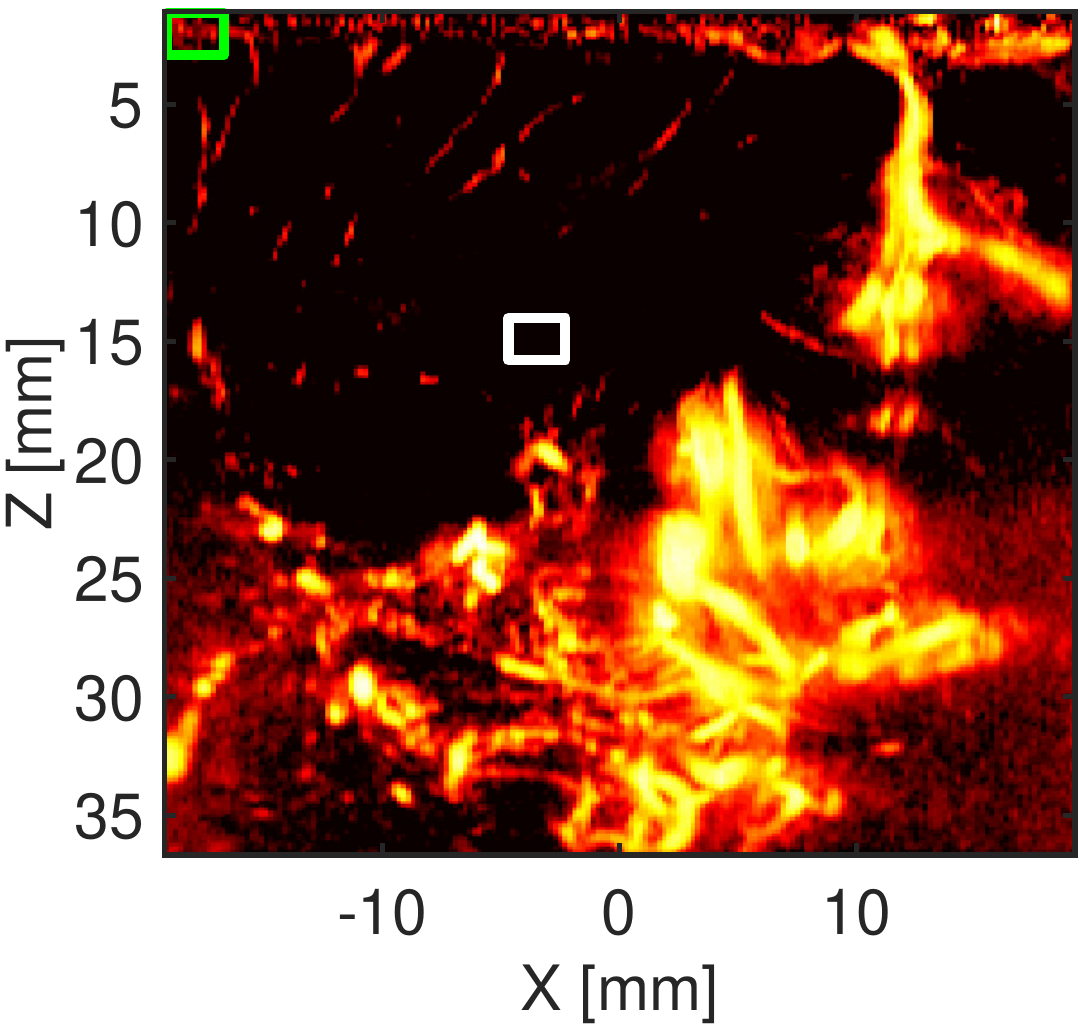}}
\centerline{{\scriptsize (h) $\lambda = 0.0049$, $\mu = 0.0098$}} 
\end{minipage}
\begin{minipage}[b]{0.05\linewidth}
\centering
\centerline{\includegraphics[trim={11.2cm 0.03cm 0 0},clip,width=0.7\textwidth, height = 6.6cm]{figures/R1/cerveau_bdrpcaold.pdf}}
\centerline{} 
\end{minipage}
\caption{ \textit{First row panels}:  Power Doppler images computed on the healthy brain dataset by respectively using: (a) SVD; (b) RPCA; (c) DRPCA; (d) BDRPCA. \textit{Second row panels}: the same as \textit{first row panel} but for the tumor dataset. All the images are displayed with a dynamic range of $35$ dB. The arrows mean that $\text{R}_2$ is successively moved patch by patch in both directions and combined with $\text{R}_1$ in order to compute CR values. The best possible hyperparameters associated with each blood recovery technique are also provided in each plot. }    
\label{fig:vivo}
\end{figure*}

\begin{figure*}[!htb]
\begin{minipage}[b]{0.48\linewidth}
\centering
\centerline{\includegraphics[width=0.85\textwidth, height = 5.5cm]{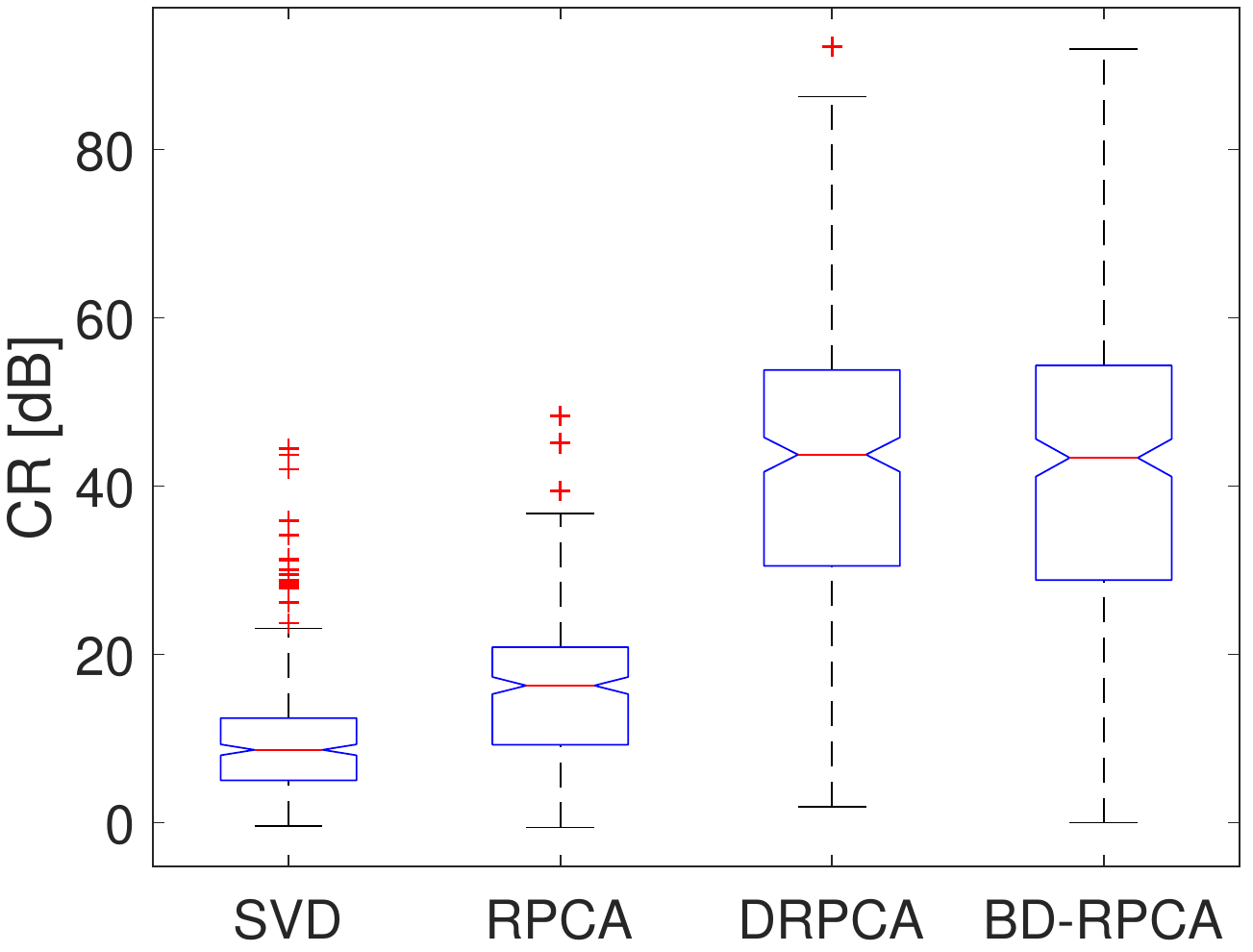}}
\centerline{{\scriptsize(a)}} 
\end{minipage}
\begin{minipage}[b]{0.48\linewidth}
\centering
\centerline{\includegraphics[width=0.85\textwidth, height = 5.5cm]{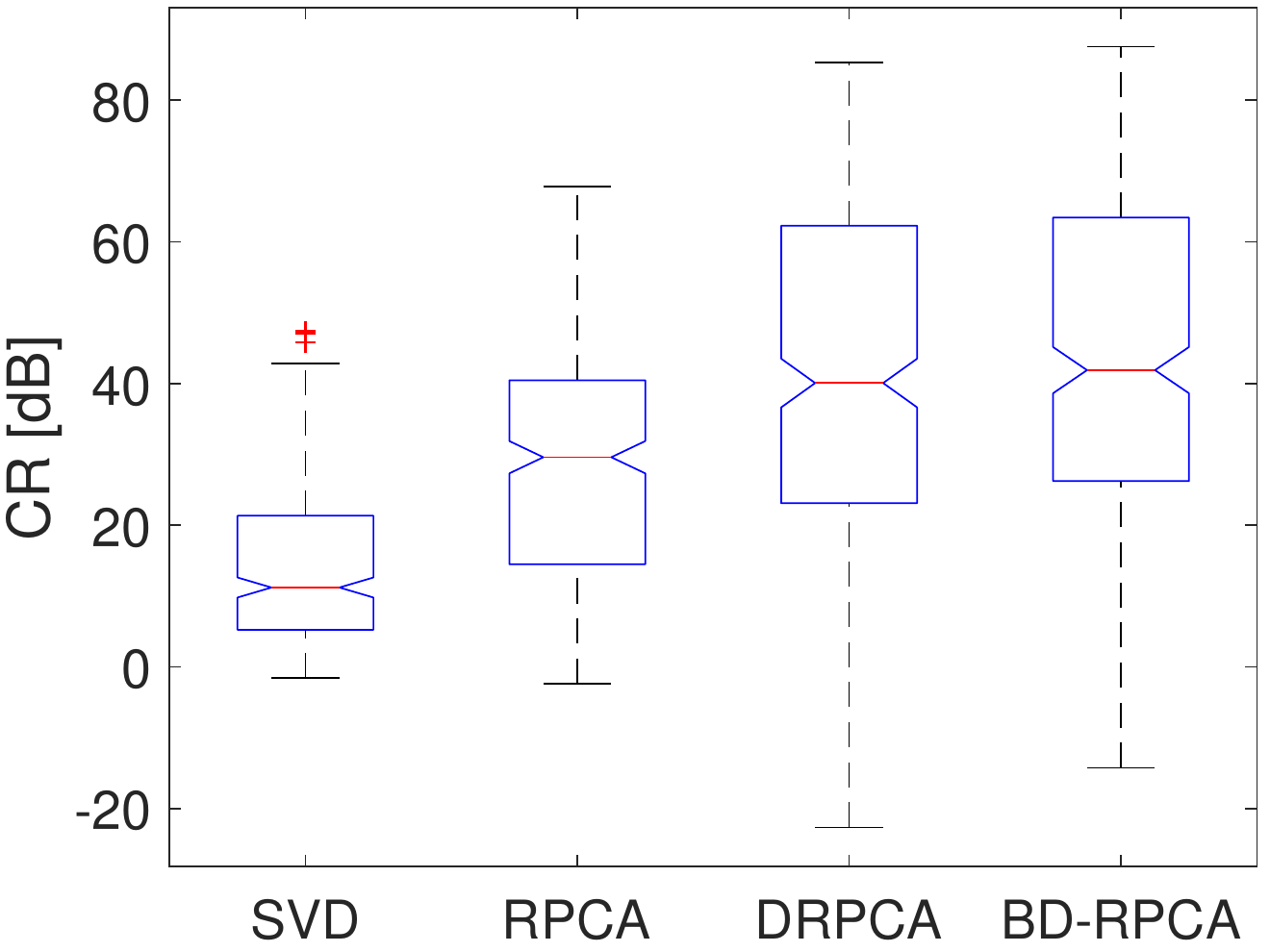}}
\centerline{{\scriptsize(b)}} 
\end{minipage}
\caption{CR measurement in dB of the different tested methods on: (a) healthy brain dataset in Fig. \ref{fig:vivo} (a); (b) tumor dataset in Fig. \ref{fig:vivo} (b). In these boxplots, the red horizontal line indicates the median, the bottom and top edges of the box indicating the 25th and 75th percentiles and the black lines indicating the entire range of data samples, per category. Red markers indicate outliers which are excluded from the statistical calculations \cite{McGill1978}.} 
\label{fig:CR}
\end{figure*}

\subsection{\textit{In vivo} results}
Ultrafast Doppler sequences were acquired on the brain of a patient undergoing a brain surgery (Regional University Hospital Bretonneaux of Tours – Department of Neurosurgery) with open skull, dura mater removed just before surgery. Two regions of interest were selected in our protocol: healthy brain with mainly large cerebral arteries and the glioma with a complex vascular \ab{structure} and very small flow in small vessels. These acquisitions were done with a clinical research protocol (ELASTOGLI) approved by the institutional review board (CCP: ‘Comité de Protection des Personnes’, CCP agreement $N\textsuperscript{{o}}$ 123748) and local ethical committee. They strictly complied with the ethical principles for medical research involving human subjects of the World Medical Association Declaration of Helsinki. \ab{Acquisitions were done using} the AixplorerTM (Supersonic Imagine) ultrasound scanner with an SL10-2 probe (192 elements). The research package (SonicLab V12) enabled to upload on the scanner a particular US sequence of $1000$ frames, compounded angles $[- 5\textsuperscript{{o}}, 0\textsuperscript{{o}}, +5\textsuperscript{{o}}]$ with pulse repetition frequency (PRF) of 3KHz, frame rate of 1KHz, imaging depth of [1mm-40mm]. One-second acquisition was downloaded on the hard disk for off-line beamforming and signal processing. The size of the two resulting datasets was $260\times192\times1000$ pixels.

In Fig.~\ref{fig:vivo}, we display Power Doppler image results given by the different reconstruction methods \ab{obtained} with the best possible hyperparameters tuned by cross-validation, on each dataset. Visually judging these plots, one may notice that both DRPCA and BD-RPCA achieve, overall, similar depiction of microvasculatures with high spatio-temporal resolution and much clearer than the two others SVD and RPCA. To assess quantitative measurements, contrast ratio (CR) introduced in \dk{\cite{RodriguezMolares2018}} was used \ab{because of the non availability} of the ground truth for the \textit{in vivo} case. The CR was computed from two rectangular patches of the same size $13 \times 12$ pixels: $\text{R}_1$ (\emph{white}) representing the background and $\text{R}_2$ (\emph{green}) representing the blood signal, taken from the same Power Doppler image as shown in Fig.~\ref{fig:vivo}. Then, CR is defined as:
\begin{equation*}
\text{CR}_{[\text{dB}]} = 20\log10 \left(\frac{\mu_{\text{R}_2}}{\mu_{\text{R}_1}}\right),
\end{equation*} 
where $\mu_{R_i}$ is the mean value of intensities in $\text{R}_i$, for $i=1, 2$. The larger the CR, the better
the performance of the blood flow estimation. Moreover, to ensure a fair evaluation, $\text{R}_1$ is kept fixed as a reference patch on the background while each Power Doppler image is divided into $320$ $13\times12$ non-overlapping patches $\text{R}_2$, which leads to $320$ CR values per each tested method. Fig.~\ref{fig:CR} shows a boxplot comparison of CR values for each of the different retrieval methods and for each dataset. From Fig.~\ref{fig:CR}, one may remark that DRPCA and BD-RPCA, overall, produce similar results and significantly better than SVD and RPCA. The quantitative results obtained by taking the median values of CR are reported in Table \ref{tab:median_realdata}. They demonstrate the \dk{consistency} with the above qualitative inspection about the performance of the blood flow estimation of the different studied techniques.     

\begin{table}[!htb]
\centering
\renewcommand{\arraystretch}{1.3}
\caption{CR median values for the \textit{in vivo} case}
\label{tab:median_realdata}
\begin{tabular}{c|c|c|c|c|}
\cline{2-5}
                            & \bfseries SVD & \bfseries RPCA & \bfseries DRPCA & \bfseries BD-RPCA \\ \hline
\multicolumn{1}{|c|}{Brain} &  8.65 & 16.29 & 43.74 & 43.36  \\ \hline
\multicolumn{1}{|c|}{Tumor}  &  21.09 & 29.60 & 42.89 & 45.21  \\ \hline
\end{tabular}
\end{table}

\begin{table}[!htb]
\centering
\renewcommand{\arraystretch}{1.3}
\caption{Running times in s for each method and each study case}
\label{tab:running_time}
\begin{tabular}{c|c|c|c|c|}
\cline{2-5}
                              & \textbf{SVD}      & \textbf{RPCA}    & \textbf{DRPCA}      & \textbf{BD-RPCA}    \\ \hline
\multicolumn{1}{|c|}{Simulation}    & 8  & 114 & 235 & 493 \\ \hline
\multicolumn{1}{|c|}{Brain}   & 13 & 21 & 116 & 212 \\ \hline
\multicolumn{1}{|c|}{Tumor} & 13 & 19 & 110 & 221 \\ \hline
\end{tabular}
\end{table}

Finally, Table \ref{tab:running_time} \ab{regroups} the running times associated with each retrieval method in both simulated and \textit{in vivo} studies. It can \dk{be} seen that the high-resolution results due to the blind deconvolution approach are at the expense of higher computational time. However, despite this limitation, both simulated and \textit{in vivo} results plead in favour of using BD-RPCA that does not make use of PSF measurement to reconstruct the blood flow rather than the other studied techniques.

\section{Conclusion}
\label{sec:concl}
In this paper, a novel algorithm for the retrieval of blood flow from an ultrafast sequence of US images was proposed, based on the combination of two different techniques DRPCA and BD. The proposed method allowed to overcome the main limitation of the former \ab{related to the requirement of PSF measurement while providing} equivalent estimation \dk{performances}. Numerical experiments demonstrated the effectiveness of the proposed technique on both simulated and \textit{in vivo} datasets. Future work will be dedicated to evaluate the clinical contribution of the proposed method, in particular its ability to improve the diagnosis power of the estimated blood flow maps. Moreover, the main drawback of the proposed algorithm is the high computational complexity; therefore, \ab{it would be of great interest to develop more computationally efficient optimization schemes to alleviate this limitation.} Finally, the PSF estimate was assumed to be spatial-temporally invariant PSF across the 3D imaging domain which constitutes a significant limitation of the proposed method; thus, taking into account the spatial-temporal variation features of the PSF as was done in \cite{Michailovich2017,Florea2018,Besson2019}, is definitely an interesting perspective.

                        
\bibliographystyle{IEEEtran}
\bibliography{Eusipco2015}
\end{document}